\documentclass[useAMS,usenatbib]{mn2e}
\usepackage{graphicx}
\usepackage{longtable}
\usepackage{url}
\usepackage{epstopdf}
\usepackage{color}
\usepackage{bm}
\usepackage{amssymb}
\usepackage{amsmath}
\usepackage{hyperref}
\usepackage{times}
\usepackage{deluxetable}
\usepackage{nccmath}

\hypersetup{
    unicode=true,                     
    linktocpage=true,
    pdftoolbar=true,                
   pdfmenubar=true,              
    pdffitwindow=true,           
    pdfstartview={FitH},        
    pdfhighlight={/I},
   hyperindex=true,
    pdftitle={My title},            
   pdfauthor={Author},         
   pdfsubject={Subject},       
   pdfcreator={Creator},       
    pdfproducer={Producer}, 
    pdfkeywords={keywords}, 
    pdfnewwindow=true,      
   colorlinks=true,       
    linkcolor=red,          
    citecolor=blue,        
    filecolor=magenta,      
    urlcolor=cyan           
}

\newcommand{\mz}{{\it M-Z}}

\newcommand{\mlimf}{$m_{\mbox{\scriptsize{lim}}}^{\mbox{\scriptsize{f}}}$}
\newcommand{\mlimb}{$m_{\mbox{\scriptsize{lim}}}^{\mbox{\scriptsize{b}}}$}

\newcommand{\mstar}{$m_*$}

\newcommand{\mstarf}{$m_*^{{\mbox{\scriptsize{f}}}}$}
\newcommand{\F}{Figure~\ref}

\newcommand{\tctv}{$T_c$ and $T_v$}
\newcommand{\mlim}{$m_{\mbox{\scriptsize{lim}}}$}
\newcommand{\tc}{$T_c$}
\newcommand{\tv}{$T_v$}
\newcommand{\vmax}{$V/V_{\max}$}
\newcommand{\sn}{({\it s/n})}

\newcommand{\deltazm}{$\delta Z$  and $\delta M$}
\newcommand{\dZ}{$\delta Z$}
\newcommand{\dM}{$\delta M$}
\newcommand{\CI}{\it Completeness I}
\newcommand{\CII}{\it Completeness II}
\newcommand{\red}{black}

\newcommand{\apjl}{ApJ Lett.}

\newcommand{\apjs}{ApJ Suppl.}

\newcommand{\aap}{A\&A}

\title[Completeness III]{Completeness III:  identifying characteristic  systematics and evolution in galaxy redshift surveys}

\author[Johnston, Teodoro \& Hendry]{Russell Johnston$^{1}\thanks{rjohnston@uwc.ac.za}$,  Lu\'{\a i}s Teodoro$^{2,\,3}\thanks{luis@astro.gla.ac.uk}$, and Martin Hendry$^2\thanks{Martin.Hendry@glasgow.ac.uk}$
\\
$^1$Department of Physics, University of Western Cape, Belville, Cape Town, South Africa\\
$^2$SUPA, School of Physics and Astronomy, Kelvin Building, University of Glasgow, Glasgow, G12 8QQ, Scotland, UK\\
$^3$BAER Int, NASA Ames Research Center, MS 245-3, Moffett Field, CA 94035-1000, USA
}

\begin{document}
\date{\today}
\pagerange{\pageref{firstpage}--\pageref{lastpage}} \pubyear{2011}

\label{firstpage}

\maketitle

\begin{abstract}
This paper continues our development of non-parametric tests for analysing the completeness in apparent magnitude of magnitude-redshift surveys. The purpose of this third and final paper in our completeness series is two-fold: firstly we explore how certain forms of incompleteness for a given flux-limited galaxy redshift survey would manifest themselves in the ROBUST {\tctv} completeness estimators introduced in our earlier papers; secondly we provide a comprehensive error propagation for these estimators.

 This work was initiated by \cite{Rauzy:2001} and then extended and developed by \citet*{Johnston:2007} ({\it Completeness I}) and \citet*{Teodoro2010MNRAS.405.1187T} ({\it Completeness II}). Here we seek to consolidate the ideas laid out in these previous papers.    In particular our goal is to provide for the observational community statistical tools that will be more easily applicable to real survey data.  By using both real surveys and Monte Carlo mock survey data, we have found distinct, characteristic behaviour of the {\tctv} estimators which identify incompleteness in the form of e.g. missing objects within a particular magnitude range.  Conversely we have identified signatures of `over' completeness, in cases where a survey contains a small region in apparent magnitude that may have too many objects relative to the rest of the data set.  Identifying regions of incompleteness (in apparent magnitude) in this way provides a powerful means to e.g. improve  weighting schemes for estimating luminosity functions, or for more accurately determining the selection function required to employ measures of galaxy clustering as a cosmological probe.

We also demonstrate how incompleteness resulting from  luminosity evolution can be identified and provide a framework for using our estimators as a robust tool for constraining models of luminosity evolution.

Finally we explore the error propagation for \tctv.  This builds on {\it Completeness II} by allowing the definition of these estimators, and their errors, via an adaptive procedure that accounts for the effects of sampling error on the observed distribution of apparent magnitude and redshift in a survey.

\end{abstract}

\section{Introduction}
The notion that we have entered an era of {\em precision cosmology\/} has been consistently cited in cosmology publications for more than a decade (see \citep{Turner:1998}). While many would support this statement when applied to CMBR measurements e.g. made by the Wilkinson Microwave Anisotropy Probe  \citep[WMAP -- see e.g.][]{Spergel:2003, Larson:2010}, in other areas, such as the study of galaxy redshift surveys, the era of precision cosmology would appear to be approaching but will require improvement in both the quality and size of our datasets and (crucially) our statistical toolbox before we can claim that it has truly arrived.

Estimating the luminosity function (LF)  remains a powerful and popular probe of galaxy evolution \citep[e.g.][]{Norberg:2002b,Blanton:2003ApJ...592..819B,Liske:2003,Richards:2006,Willmer:2006ApJ...647..853W,Faber:2007,2007MNRAS.381.1548K, Bouwens:2008,Siana:2008,Crawford:2009,Brusa:2009ApJ...693....8B,Zucca:2009,Croom:2009,Haberzettl:2009,Montero:2009MNRAS.399.1106M,Willott:2010,2010MNRAS.406.1944W,Aird:2010MNRAS.401.2531A,Rodighiero:2010AA...515A...8R}. However, its accurate determination remains one of the most fundamental statistical challenges in modern observational cosmology.

The methodology employed to estimate the LF varies greatly throughout the literature. Some of these approaches have been non-parametric, i.e. adopting no specific parametric model for the LF, and have ranged historically from the classical number count test \citep[e.g][]{Hubble:1936,Christensen:1975}, the \cite{Schmidt:1968} $1/V_{\max}$ estimator, the $\phi/\Phi$ method \citep[e.g.][]{Turner:1979}, and the \cite{lynden:1971} $C^-$ method.  Alternatively, there have also been several parametric methods developed, generally based on the Maximum Likelihood Estimator (MLE) method of \citep{sandage:1979} (STY), where a parametric form  of the LF  \citep[most commonly that of][]{Schechter:1976} is assumed. Thirdly,  there exists a `hybrid' method: the non-parametric counterpart of the MLE developed by \cite{Efstathiou:1988}, and often referred to as the Stepwise Maximum Likelihood method (SWML).   Of course we have not listed the many variations of these three broad approaches that have arisen as a result of a now thriving industry that has produced (and continues to produce) the myriad of complex and tailor-made galaxy redshift surveys.  For more detailed articles that trace the origins and development of this area of statistics the reader is referred to  e.g. \cite{2011AARv..19...41J,Binggeli:1988,Willmer:1997} and \cite{Takeuchi:2000}

Recently the field of LF estimation has witnessed some significant fresh developments, where new approaches have emerged with the goal of placing the methodology on a more rigorous statistical footing. Examples of these include: the semi-parametric  approach  by \cite{Schafer:2007}; Bayesian methods of   \cite{Andreon:2006MNRAS.369..969A} and \cite{Kelly:2008},  the use of the  {\it copula}\footnote{The copula  is a function used to  {\it join} multivariate distribution functions to their one-dimensional marginal distribution function and is particularly useful for variables with co-dependence.}   by  \cite{Takeuchi2010MNRAS.406.1830T} to   construct the bivariate LF; and a non-parametric inversion technique by \cite{Borgne:2009AA...504..727L} applied  to galaxy counts.

Two key  fundamental assumptions that are common to almost all LF estimators and which are also crucial to the work detailed in this paper can be summarised as:
\begin{enumerate}
\item{{\it Separability} between the luminosity function, $\phi (M)$,  and the density function, $\rho(z)$, probability densities; i.e. one assumes that the underlying joint distribution of luminosity (equivalently absolute magnitude) and redshift may be written as the product of their marginal distributions.}
\item{{\it Completeness} in apparent magnitude up to a specified faint apparent magnitude limit or within specified bright and faint magnitude limits.}
\end{enumerate}
For clarity, note that we define completeness in this context as the probability that a galaxy of apparent magnitude, $m$, is observable.

In \cite{Johnston:2007}  (hereafter  {\CI}) and \cite{Teodoro2010MNRAS.405.1187T}  (hereafter {\CII}) we discussed in some depth the relative merits of traditional apparent magnitude completeness tests -- e.g. the classical number count test \citep{Hubble:1926} and the still widely used \cite{Schmidt:1968} {\vmax} test.  In particular we highlighted that both tests are susceptible to bias when applied to a survey which is spatially inhomogeneous.  More specifically, we noted that in practice it can be difficult to decide whether deviations from the expected value of the respective test statistics are indeed the result of survey incompleteness in apparent magnitude, or are an artefact of galaxy clustering and/or evolution of the galaxy luminosity function.   However, at least in the case of {\vmax} (and its $1/V_{\max}$ counterpart), this issue has  not diminished the widespread application and extension of these tests -- possibly due to their simple implementation \citep[see .e.g.][]{Huchra:1973,Felten:1976,avni:1980,Hudson:1991,Eales:1993,Qin:1997,Qin:1999,Page:2000,Sheth:2007}.  Nevertheless, we believe that this should not deter us from developing better statistical tests that are less prone to such biases when the assumptions underpinning them are not fully satisfied.

\cite{Efron:1992} (hereafter EP92) revisited the properties of the  seminal \cite{lynden:1971}  $C^-$ method for constructing galaxy LFs.  Their paper introduced a powerful new approach to analysing magnitude-redshift surveys in which they proposed a non-parametric permutation test of the independence of the spatial and luminosity distributions of galaxies in a magnitude-limited sample.  As with the $C^-$ method, the EP92 test  required no assumptions concerning the parametric form of both the spatial distribution and the galaxy luminosity function.  However, this estimator  assumes the data under test {\it is\/} complete in apparent magnitude. They applied their permutation test to a quasar sample, with an assumed apparent magnitude limit, in order to robustly estimate the parameters characterising the luminosity distance-redshift relation of the quasars \citep[see also][]{Efron:1998}.  These permutation test tools have since been adopted to explore e.g. evolutionary  models for the dependence of Gamma Ray Burster (GRB) luminosity with redshift \citep{Lloyd:2002}, correlations between the luminosity of galactic nuclei and that of their host galaxy \citep{Hao:2005}, and, more recently, the radio and optical luminosity functions in \cite{Singal:2011arXiv1101.2930S}.

\cite{Rauzy:2001} (hereafter R01) extended the ideas of EP92  by adapting them to develop a simple but powerful tool for assessing the apparent magnitude completeness of magnitude-redshift surveys. As was the case with EP92 -- and unlike the Hubble number counts or {\vmax} tests -- the Rauzy test statistic {\tc} requires {\it no\/} assumption that the spatial distribution of galaxies is homogenous.  Moreover, it also requires no assumption of a specific parametric form for the galaxy luminosity function. However, the Rauzy test was formulated only for the case of an assumed sharp, faint apparent magnitude limit. The Rauzy test has since been applied to a wide variety of data exploring e.g.  the completeness limits of the HI mass function in the HIPASS survey  \citep{Zwaan:2004MNRAS.350.1210Z},  the HI flux completeness of ALFALFA survey data \citep{Toribio:2011arXiv1103.0990T}, for nearby galaxies dwarf galaxies in the local volume \citep{Lee:2009}.  A recent study by \cite{Devereux:2009} adopted the Rauzy method to determine the completeness of multi-wavelength selected data.

{\CI} extended the Rauzy test to the more realistic case of data characterised by both a faint {\it and\/} bright apparent magnitude limit.  Moreover we introduced a new variant statistic, denoted {\tv}, that samples the cumulative distance modulus, $Z$, distribution but retains similar robust properties to those of {\tc} -- i.e. being independent of the spatial distribution of galaxies.  By sampling the data in this way, the {\tv} statistic can be considered as an improved, differential version of the the widely used {\vmax} test (which assumes spatial homogeneity).

Our next paper, {\CII}, highlighted the fact that the previously defined completeness estimators may be susceptible to `shot-noise' effects if studying completeness in a small parent data-set.  This phenomenon can be explained in the context of {\CI}, where the basic construction of the  {\tctv} statistics allocated a volume limited subsample to each individual galaxy in the catalogue. With the introduction of a secondary bright apparent magnitude limit (as illustrated in \F{Fig:0MZ}) the size of the volume limited subsample constructed for each galaxy may be significantly restricted. {\CII} showed how this may render the \tctv\,completeness estimators `shot-noise' dominated when the volume limited regions are very sparsely sampled.  As a result {\CII} proposed an extension of the basic \tctv completeness estimators facilitating their construction in a manner that essentially maintains a constant `signal-to-noise' ({\sn}) level.

This article examines more deeply characteristic forms of incompleteness and how they will be manifested in the resulting {\tctv} estimators.  We also  further extend the tools developed in {\CII}, where we now introduce an error propagation analysis for the {\tctv} statistics. Hence we continue our task of providing a statistically rigorous, but practical, foundation for testing the magnitude completeness of current and future redshift surveys.

The structure of this paper is as follows.  In \S~\ref{sec:statsframework} we briefly outline the basic framework that has underpinned the development of the {\tctv} statistics, from R01 through to the the signal-to-noise approach detailed in {\CII}. In \S~\ref{sec:character} we explore various forms of incompleteness   through the use of real and simulated data. In \S~\ref{sec:errorprop} we derive expressions for the error propagation of {\tctv}.  Finally in \S~\ref{sec:conclusions} we discuss our findings and summarise our conclusions.
\section{Building the {\it complete} picture}\label{sec:statsframework}
\begin{figure*}
     		\includegraphics[width=0.49\textwidth]{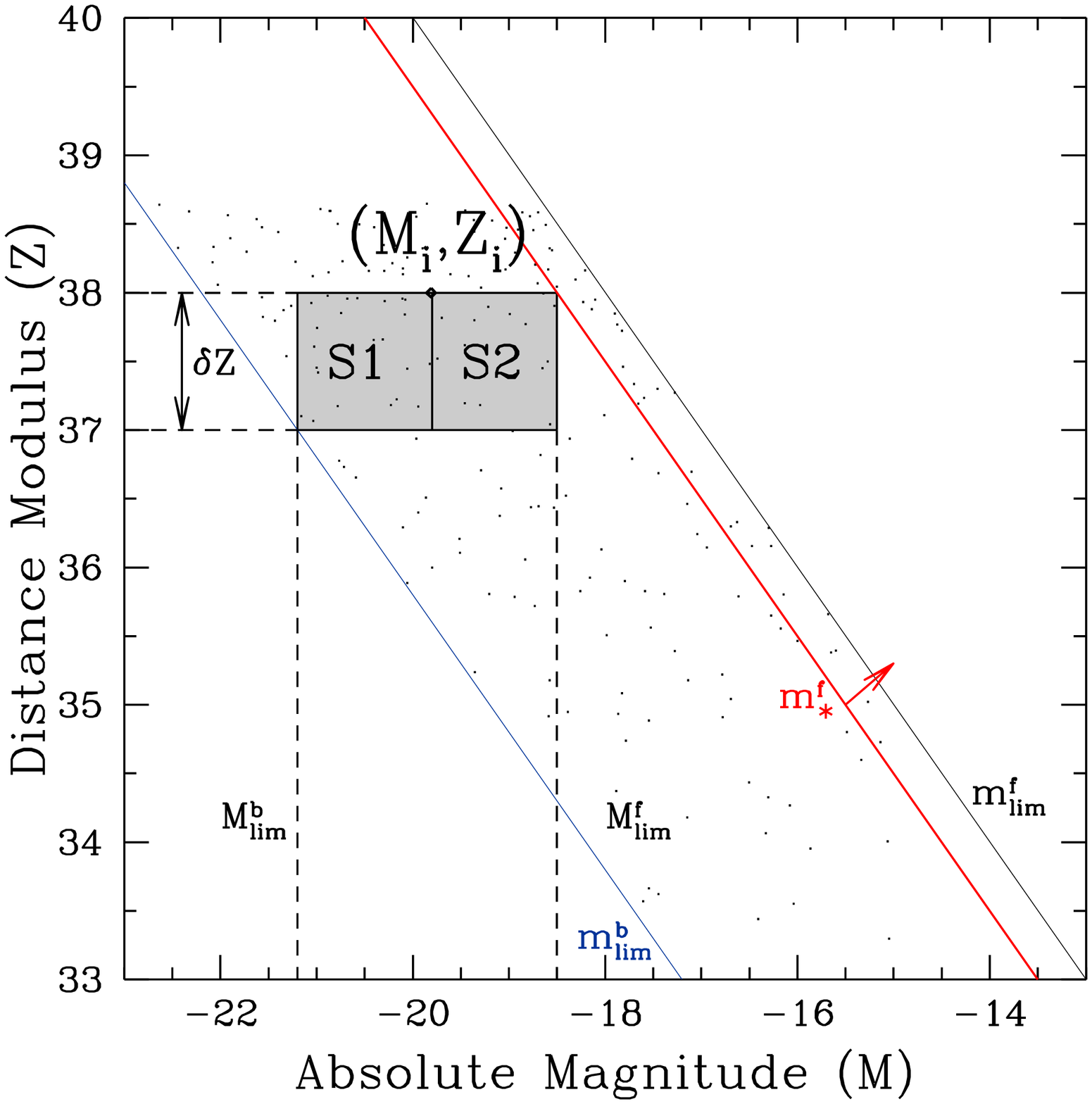}
   		\includegraphics[width=0.49\textwidth]{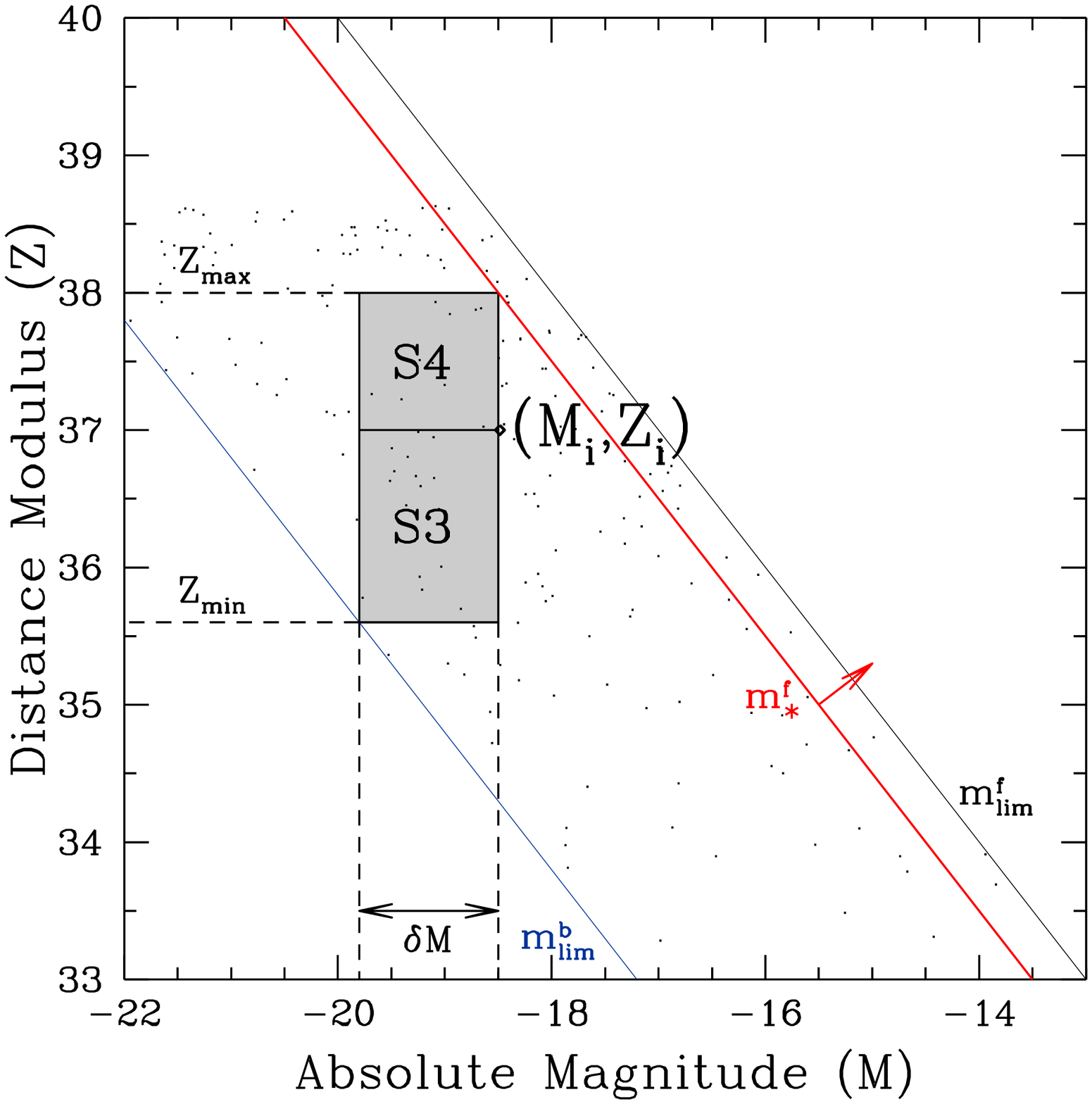}
     \caption{\small Diagram illustrating the construction of the rectangular regions used to define the random variables, $\zeta_i$(\mstarf) (left panel) and $\tau_i$(\mstarf) (right panel), for a typical galaxy at $(M_i,Z_i)$ drawn from a survey that is subject to bright and faint apparent magnitude limits {\mlimb}  and {\mlimf} respectively. The left hand panel illustrates the construction of the regions $S_1$ and $S_2$, related to the random variable $\zeta_i$, for a given `trial' value \mstarf of the faint apparent magnitude limit.  These regions are uniquely defined for a rectangular slice of fixed width, {\dZ}, in distance modulus.  The right hand panel illustrates the construction of the corresponding rectangular regions $S_3$ and $S_4$, related to the random variable $\tau_i$, for a given trial value of (\mstarf). Similarly these regions are uniquely defined for a rectangular `slice' of width, {\dM}, in absolute magnitude. See text for further details.
      }
      \label{Fig:0MZ}
  \end{figure*}
In this section we briefly outline all the elements essential to defining  our completeness estimators, as previously developed in \cite{Rauzy:2001} (R01), \cite{Johnston:2007} ({\CI}) and \cite{Teodoro2010MNRAS.405.1187T} ({\CII}).  The reader is referred to those articles for a more detailed discussion.
\subsection{The assumption of separability}
In R01 the foundations of our statistical method were established on the assumption of separability, whereby  the luminosity function of the galaxy distribution is considered to be independent of the the three-dimensional redshift space coordinates {\it{\bf{z}}} $= (z,l,b)$ of the surveyed galaxies, (where $(l,b)$ are galactic coordinates).  As we have already discussed in our previous papers, this is a rather restrictive assumption which, nevertheless, underpins  most of the traditional completeness tests (and indeed most luminosity function estimators as well) in the literature.  In future work we will investigate methods to further adapt our completeness estimators, exploiting departures from the null hypothesis of separability as a sensitive probe of galaxy evolution.  For the remainder of this paper, however, we will generally restrict our attention to cases where the assumption of separability is satisfied.
\subsection{Defining the fundamental variables, $Z$ and $M$}
We begin by considering the {\it uncorrected\/} distance modulus $Z$.  By uncorrected we mean (in the first instance) that  $K$- and source evolution ($e$-) correction terms have been neglected.   Although this approach differs from that of R01, where such corrections were integral to the formulation in that paper, our subsequent experience of applying our completeness estimators has demonstrated that it is more instructive to begin with the {\it raw\/}  (i.e. only initially corrected for extinction) magnitudes; one can then add e.g. K- and evolution correction terms in an incremental fashion, thus more readily determining the impact of each correction on the magnitude completeness of the survey.  Therefore, $Z$, is simply defined as,
\begin{equation}
Z=5\log_{10}(d_L)+25 \equiv m-M,
\end{equation}
where $Z$ is the distance modulus corresponding to redshift $z$, $m$ is  apparent magnitude corrected for extinction only and $d_L$ is the luminosity distance.   The absolute magnitude, $M$,  is then equivalently defined as,
\begin{equation}
M=m-5\log_{10}(d_L) - 25 
\end{equation}
Note also that in what follows, for simplicity, we will suppress explicit reference to the angular coordinates and work with the distribution of $Z$ marginalised over $l$ and $b$.
\subsection{The random variables, $\zeta$(\mstarf) and $\tau$(\mstarf)}
The {\tctv} statistics are constructed from estimates of the random variables, $\zeta$(\mstarf) and $\tau$(\mstarf) respectively (see  {\F{Fig:0MZ}}) which are defined in terms of the joint probability density in $M$ and $Z$, constructed in a  separable form. {\mstarf} represents the trial faint apparent magnitude limit under test indicated by the red diagonal lines in {\F{Fig:0MZ}}.  Thus, following R01 and {\CI}, for each object $i$ present in a catalogue we define
\begin{equation}\label{equ:tc}
T_c(\mbox{\mstarf}) = \frac{\displaystyle \sum_{i=1}^{N_{\rm gal}} \left [ \zeta_i(\mbox{\mstarf})-1/2 \right] }{\displaystyle\left[\sum_{i=1}^{N_{\rm gal}}\mbox{Var}(\zeta_i)\right]^{1/2}},
\end{equation}
and
\begin{eqnarray}
T_v(\mbox{\mstarf})= \frac{\displaystyle\sum_{i=1}^{N_{\rm gal}} \left[ \tau_i(\mbox{\mstarf})-1/2 \right]}{\displaystyle \left[\sum_{i=1}^{N_{\rm gal}}\mbox{Var}(\tau_i)\right]^{1/2}}, 
\end{eqnarray}
where
\begin{eqnarray}\label{Eq:zetav1}
\zeta_i (\mbox{\mstarf})&=& \frac{F(M_i) - F [ M^{\rm b}_{\rm lim}(Z_i - \delta Z) ]}{F [M^{\rm f}_{\rm lim}(Z_i) ] - F [ M^{\rm b}_{\rm lim}(Z_i - \delta Z) ]} \nonumber \\
&= & \frac{n({S_1})}{n({S_1\cup S_2})}=\frac{r_i(\mbox{\mstarf})}{n_i(\mbox{\mstarf})+1},
\end{eqnarray}
and
\begin{eqnarray}\label{Eq:tauv1}
\tau_i(\mbox{\mstarf}) &=&\frac{H(Z_i) - H[Z^{\mbox{\scriptsize{b}}}_{\rm min}(M_i-\delta M)]}{H[Z^{\mbox{\scriptsize{f}}}_{\rm max}(M_i)] - H[Z^{\mbox{\scriptsize{b}}}_{\rm min}(M_i-\delta M)]} \nonumber \\ & =&  \frac{n({S_3})}{n({S_3\cup S_4})}=\frac{q_i(\mbox{\mstarf})}{t_i(\mbox{\mstarf})+1},
\end{eqnarray}
and $r_i$ denotes the number of galaxies belonging to region $S_1$, $n_i$ the number of galaxies belonging to $S_1\,\cup\,S_2$, $q_i$ the number of galaxies belonging to $S_3$, and  $t_i$ the number of galaxies belonging to $S_3\,\cup\,S_4$. $F(M)$ is the cumulative luminosity function and  $H(Z)$ is the cumulative distribution function. The variance, ${\rm Var}(\zeta_i)$ and ${\rm Var}(\tau_i)$, are respectively given by,
\begin{equation}\label{equ:var}
{\rm Var}(\zeta_i)=\frac{1}{12}\frac{n_i-1}{n_i+1}
\end{equation}
and
\begin{equation}\label{equ:var}
{\rm Var}(\tau_i)=\frac{1}{12}\frac{t_i-1}{t_i+1}
\end{equation}
 Figure~\ref{Fig:0MZ} illustrates the construction of the rectangular regions $S_1$, $S_2$, $S_3$ and $S_4$ as well as the meaning and definition of the slices in magnitude, {\dZ}, and distance modulus, $\delta M$. It should be mentioned that $r_i$ was also the notation used in EP92 to denote the {\it rank} of  the object $i$ when galaxies are sorted by magnitude. In the scenario illustrated in Figure~\ref{Fig:0MZ} the quantities are {\dZ} and {\dM} are fixed quantities. Therefore, one can see that initial {\mstar} values close to the bright limit, {\mlimb},  where a chosen value of {\dZ} and {\dM} is too large for sufficient sampling, neither  {\tc} or {\tv}  will be calculated.  Moreover,  as {\mstar} moves to increasingly fainter magnitudes, any objects too \textcolor{\red}{ close } to {\mlimb} for a $\zeta_i$({\dZ}) or $\tau_i$({\dM}) calculation to be made will be omitted from the final completeness calculation.
 \subsection{Signal-to-noise \& shot-noise sampling}
In {\CI} we identified two effects that are a consequence of adopting fixed `slice' widths {\dZ} for $\zeta_i$ and {\dM} for $\tau_i$.  Note that fixing these widths to predetermined values allows the construction of unique, separable regions, following Equations~{\ref{Eq:zetav1} and \ref{Eq:tauv1}, within any survey with a well defined bright and faint apparent magnitude limit (usually referred to as {\em doubly truncated\/}).
However {\CII} explored the sensitivity of our results to the (essentially arbitrary) choices of {\deltazm}. It was identified that,
\begin{enumerate}
\item For  very  small values of  {\dZ} and {\dM} the respective {\tctv} statistics will be dominated by what we may term `shot-noise' (since the rectangular regions they identify are extremely sparsely sampled); this makes the process of drawing significant conclusions regarding the nature of the true faint apparent magnitude limit impossible.
\item{Conversely, when the values of  {\deltazm} are taken to be very large, then for data-sets that are not well described by a single, sharp faint end apparent magnitude limit {\mlim} one observes that the behaviour of the {\tctv} statistics appears to be sensitive to values of {\deltazm} -- which may lead to inconsistent conclusions about the {\em true\/} faint magnitude limit.}
\end{enumerate}
These effects prompted us in {\CII} to derive expressions for an effective signal-to-noise ratio {\sn} for {\tctv}, given respectively by,
\begin{equation}\label{Eq:zetasigtonoise}
\frac{{\zeta_i }}{{(\delta \zeta_i ) }}
= \left [ \frac{{r_i^2 }}{{\delta r_i ^2 }}
+ \frac{{(n_i   + 1)^2 }}{{[\delta (n_i  + 1)]^2 }}
- \frac{{r_i(n_i   + 1)}}{{2\delta r_i [\delta (n_i   + 1)]}}\right ]^{1/2},
\end{equation}
and,
\begin{equation}\label{Eq:tausigtonoise}
\frac{{\tau_i}}{{(\delta \tau_i)}}
 = \left [ \frac{{q_i^2 }}{{\delta q_i ^2 }}
 + \frac{{(t_i   + 1)^2 }}{{[\delta (t_i  + 1)]^2 }}
 - \frac{{q_i(t_i   + 1)}}{{2\delta q_i [\delta (t_i   + 1)]}}\right ]^{1/2},
\end{equation}%
One can use these expressions to motivate a different method for choosing the widths {\deltazm}, one that resembles an ``adaptive smoothing" procedure: in short we allow the values of {\deltazm} to vary from galaxy to galaxy, so as to maintain a constant signal-to-noise ratio for the statistics {\tctv} -- thereby seeking to maintain the same amount of  information, as measured by the signal-to-noise ratio, allocated to each galaxy within the separable regions that define our estimators of $\zeta$ and $\tau$ respectively.  The appropriate minimum target signal-to-noise value was determined by trial and error, based on the  {\deltazm} values for which the {\tctv} statistics would fall below the $-3\sigma$ level at the true apparent magnitude limit.

\section{characterising incompleteness}\label{sec:character}
The key  motivation for the original Rauzy completeness test, was to provide a simple non-parametric method to determine the true magnitude limit of a magnitude limited survey. The R01 and subsequent {\CI} and {\CII} papers demonstrated effectively how this is achieved.  We now extend the use of our estimators beyond probing just the faint limit of a survey.  This section firstly demonstrates the robustness of our estimators before going onto to explore the different  forms of systematics to which {\tctv} should be particularly sensitive, noting the characteristic features  of the {\tctv} curves that are the result of incomplete patches in a given survey or the signatures of luminosity evolution.
\subsection{Probing the  robustness of {\tctv}}
As we have already discussed, our completeness estimators assume separability between the luminosity function $\Phi(M)$ and the density function $\rho(Z)$.  Therefore, the estimators should be sensitive only to systematic departures from this assumption and conversely they should be essentially {\it insensitive\/} to random departures from separability. In this section we demonstrate that this is indeed the case.
\subsubsection{Randomly removing sources}
\begin{figure}
	\begin{center}
     		 \includegraphics[width=0.49\textwidth]{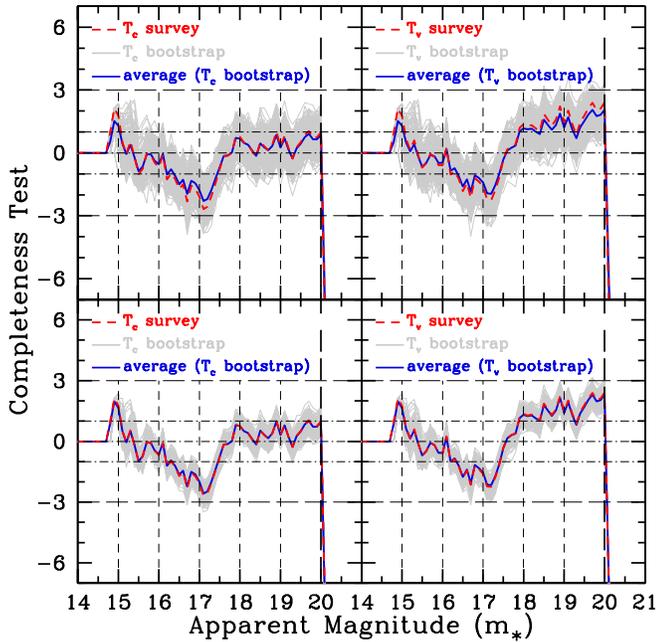}\hfill
     \caption{\small The top panels show the resulting bootstrap analysis when, for each bootstrap sample, we uniformly randomly remove 2000 sources from the parent catalogue. Similarly the bottom panels show the resulting {\tctv} curves when we randomly remove 500 sources from the parent catalogue.  On each panel is shown completeness results for the actual MGC survey (red dashed line), The bootstrapped samples (light grey lines), and the averaged bootstrapped samples (solid blue line).}
     
      \label{Fig:ranremove}
     \end{center}
  \end{figure}
First we  introduce a random sampling of the underlying luminosity function.   As with our previous studies in {\CII} and {\CII} we use the Millennium Galaxy Catalogue (MGC) to perform this  analysis (for full details of the sample selection the reader is referred to those papers). To summarise, the sample contains 7878 galaxies out to a limiting magnitude of $m=20.0$~mag in a redshift range $0.015 < z < 0.18$. By applying  a simple bootstrap-by-replacement algorithm we created 2 sets of a total of 500 realisations of the parent catalogue, randomly removing  500 then 2000 sources from each realisation.  The resulting {\tctv} curves are shown in  Figure~\ref{Fig:ranremove}.  On each panel the completeness test statistics for the full survey are shown in red. The grey lines represent the test statistics for each bootstrap sample and the blue line shows the average of these test statistics.  From the bottom panels we can clearly see that when one removes 500 sources at random the resulting {\tc} (left) and {\tv} (right) curves continue to trace the full survey data extremely well.  In fact, the averaged line in blue almost completely overlaps the test statistic for the full survey data, showing no sign of any systematic departure from the original survey completeness result.  In the top panels we extend this to randomly removing 2000 objects in each bootstrap sample, i.e. approximately a quarter of the parent dataset. Both {\tctv} statistics show a broader distribution compared to the top panels, but overall the averaged values deviate only slightly from the original survey data.  From this  simple test we  conclude that when the source removal  is performed randomly the $T_c$ and $T_v$ estimators remain robust.
\subsubsection{Retaining Separability}
We can also demonstrate the insensitivity of {\tctv} to the spatial distribution and luminosity function of the sampled objects. To do this we remove objects from slices of distance modulus ($Z$) and absolute magnitude ($M$) respectively. Figure~\ref{Fig:verthoriz} illustrates these two scenarios.

\begin{figure*}
	\begin{center}
	  \includegraphics[trim=0 0 0 3cm, width=0.5\textwidth]{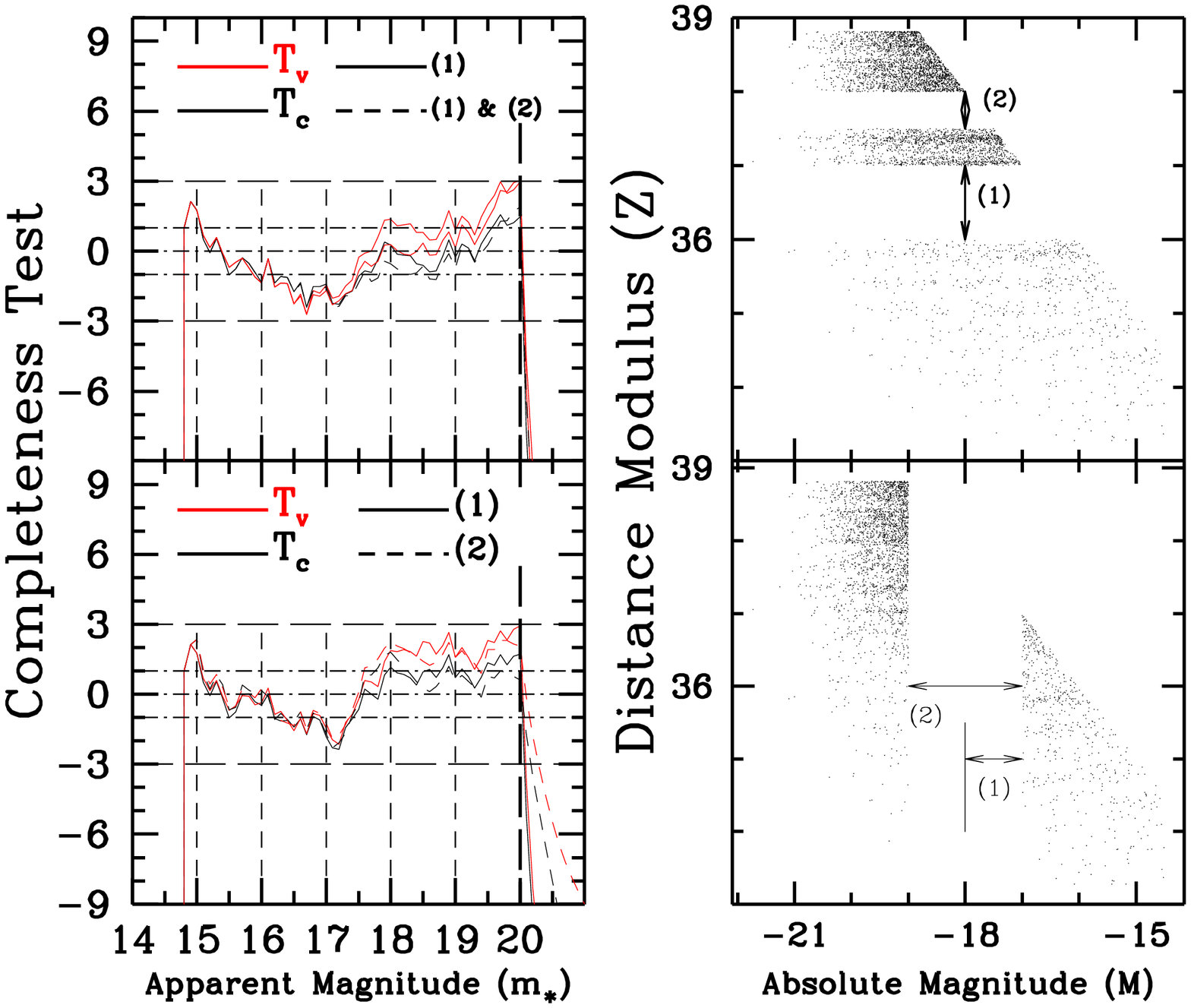}\hfill
	   \includegraphics[trim=0 0 0 3cm, width=0.5\textwidth]{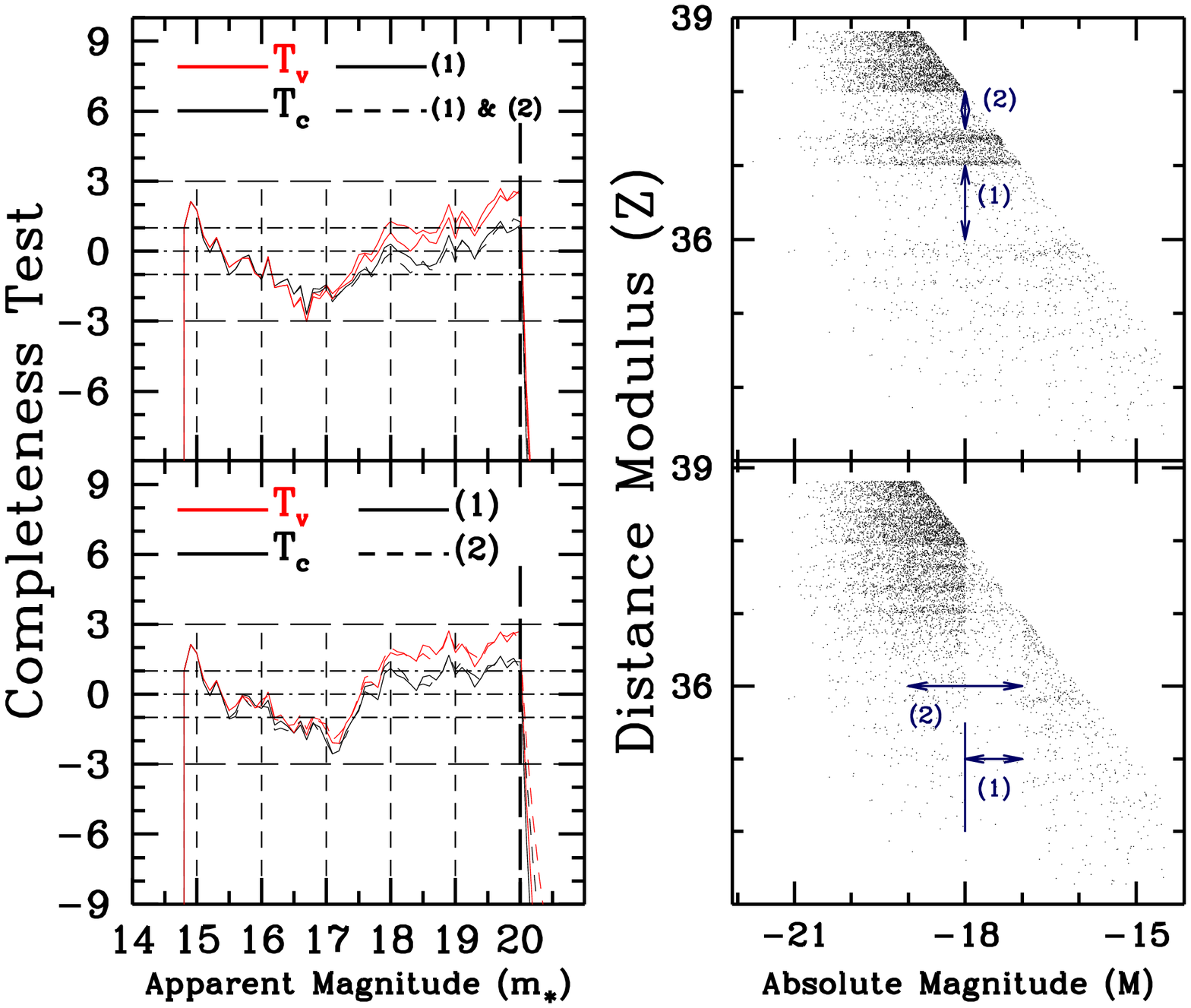}

	      \caption{\small Retaining a separable distribution. Using the MGC data we can demonstrate the robustness of {\tctv} to the removal of regions on the {\mz} plane, in two distinct cases where a separable distribution is retained. The four panels on the left show that removing either horizontal or vertical sections of the {\mz} distribution does not adversely affect the {\tctv} outcomes. Furthermore, the four right hand panels show that if we then randomly replace approximately 80\% of the objects within the removed regions, the statistics remain largely unaffected.  This is a direct result of the construction of both {\tctv} where they remain independent of any clustering effects.  It is as a consequence of this robustness that the estimators are conversely very sensitive to systematic changes in the distribution of {\it apparent\/} magnitude (see Figure~\ref{Fig:sys2}). }
	      \label{Fig:verthoriz}
\end{center}
  \end{figure*}
\begin{figure*}
	\begin{center}
	  	\includegraphics[trim=0 0 0 3cm, width=0.5\textwidth]{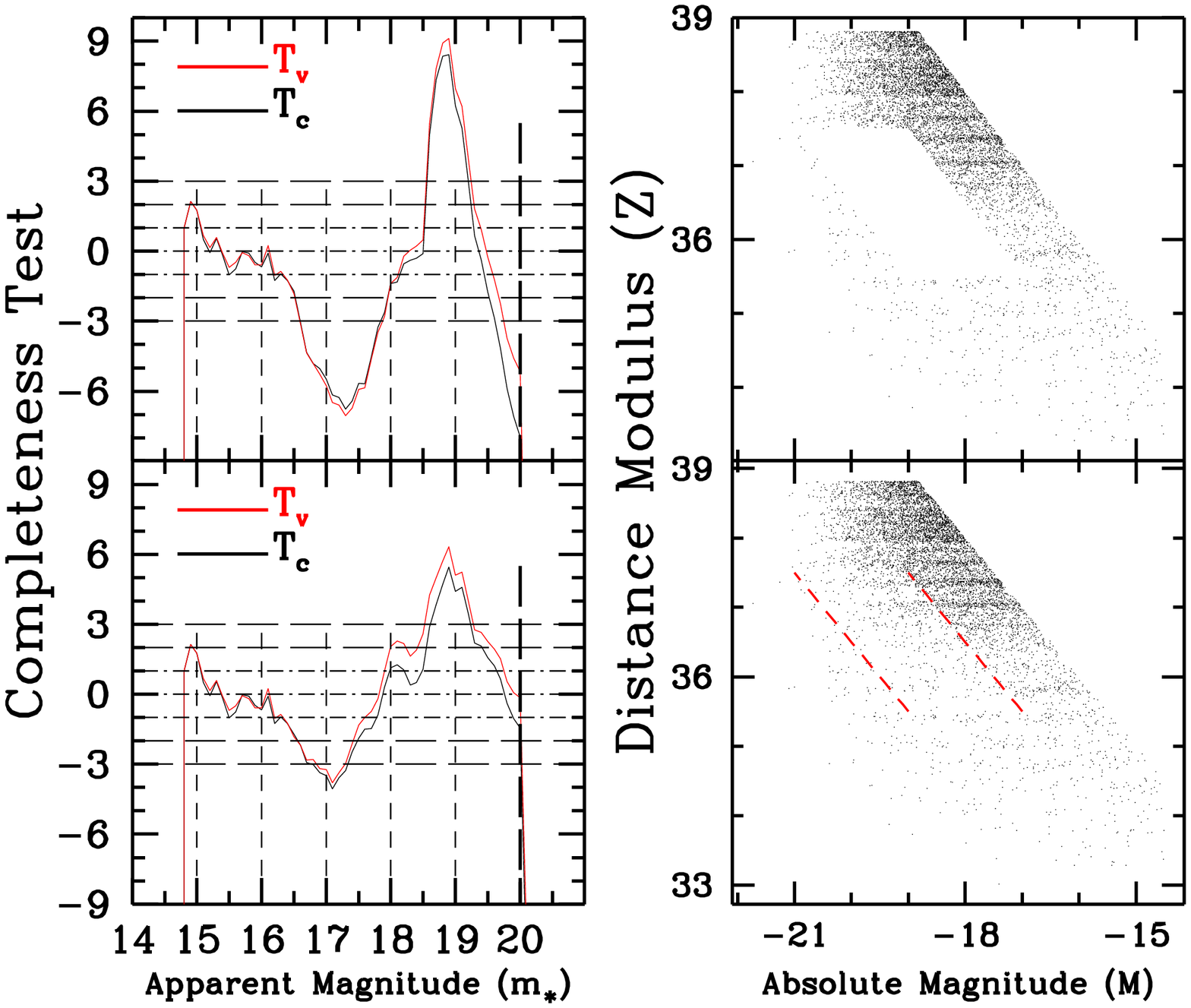}\hfill
	  	 \includegraphics[trim=0 0 0 3cm,  width=0.5\textwidth]{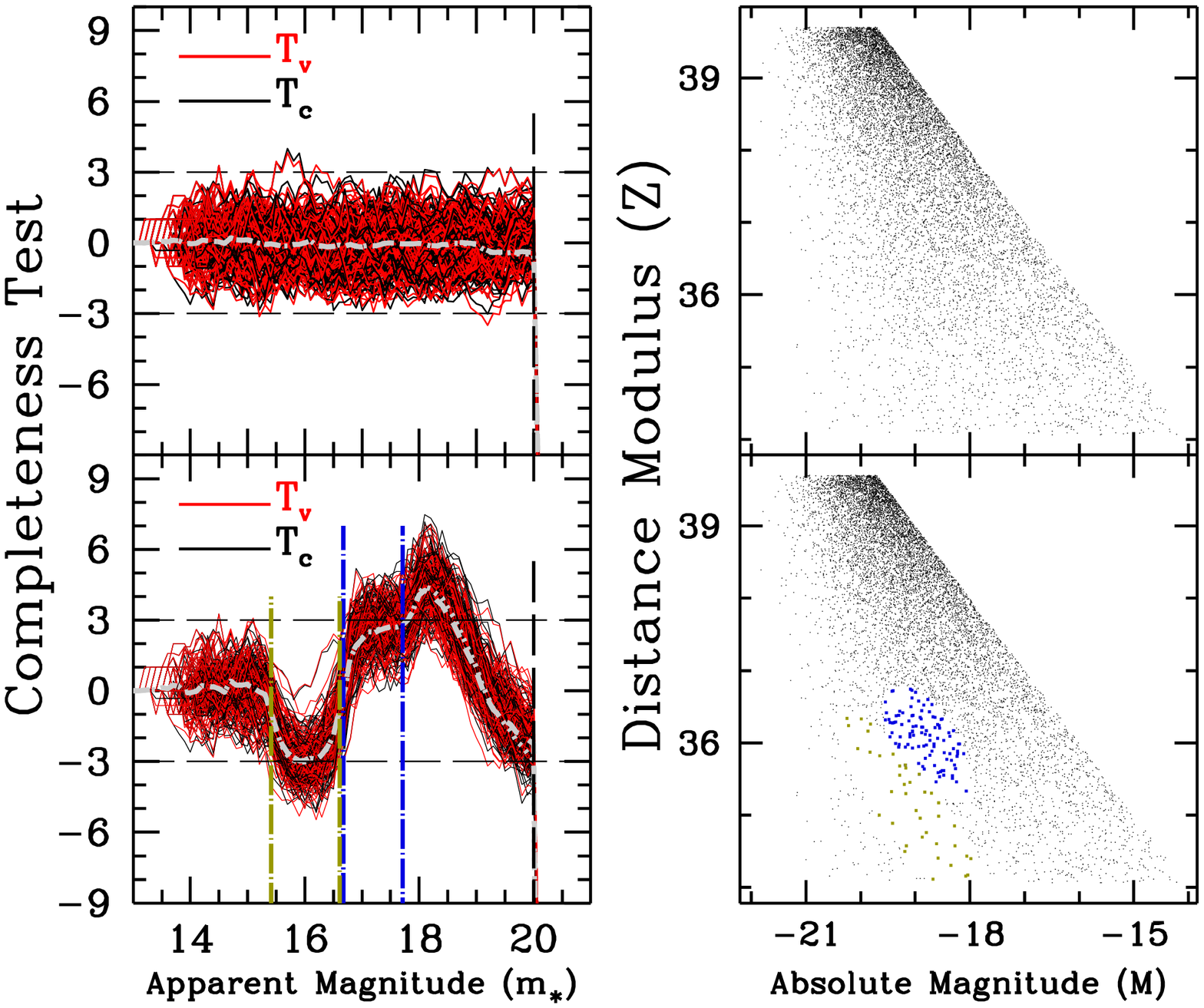}
	      \caption{\small Introducing crude systematic effects: 
missing objects. In the top left panels we have completely removed objects in the range $16.5 <m<18.5$ and $35.5<Z<37.5$ from the original MGC survey data.  The corresponding {\tctv} statistics are shown in black and red respectively. In the bottom left panels we now randomly return approximately half of the galaxies that were removed.  The red dashed line indicates the boundaries in apparent magnitude of this region. Whilst, `by eye' the {\mz} distribution in the bottom left panel may appear complete again, the {\tctv} statistics are sensitive enough to detect the residual incompleteness resulting from the fact that only about half of the removed galaxies were reinstated. The top right  panels show the superimposed {\tctv} estimator curves for 100 mock catalogues, with the {\mz} distribution for one of these mock catalogues shown in the rightmost panel.  These mock catalogues were drawn from a Universal Schechter function and as such should show no signs of incompleteness. \textcolor{\red}{This is confirmed in the top right panels where grey dashed and dotted  lines indicate the averaged values for {\tctv} respectively (it should be noted that since both {\tctv} both trace each other very closely, distinguishing between their averaged values on the plot is difficult)}.  In the bottom right panels we now take these mock catalogues and uniformly randomly remove from them some of the galaxies in 2 small ellipsoidal regions, as indicated by the two colours, blue and green of the sources that remain on the right-hand panel. This crude procedure amounts to the removal of approximately only 4\% of each parent catalogue, but already we can see {\tctv} are sufficiently sensitive that they show a characteristic, strong deviation over the relevant range of apparent magnitudes.  The average of the {\tctv} statistics in this case displays a similar structure to that of each of the mock catalogues.}
	      \label{Fig:sys2}
\end{center}
  \end{figure*}
By their construction both {\tctv} should be insensitive to spatial inhomogeneities in redshift (or distance modulus).  For illustrative purposes we demonstrate this robustness in an extreme way, as shown in the top panels of Figure~\ref{Fig:verthoriz}. The top-left panel set shows two cases where strips of galaxies in $Z$ have been completely removed and the corresponding completeness has been re-assessed. In case (1) in the top left panel set, all galaxies have been removed between $36<Z<37$. The resulting {\tctv} completeness results show little deviation from the completeness results for the parent data-set. We then remove a second strip at $37.5<Z<38$. The resulting {\tctv} (dashed lines) statistics are once again not adversely affected.  On the right-hand panel set we now randomly replace half the galaxies that were removed from each strip to observe if any bias in the statistics is introduced. As is evident from the corresponding completeness curves, there is very little perceptible change.

In a similar way we can instead remove galaxies in strips of absolute magnitude, as we show in the bottom panels of  Figure~\ref{Fig:verthoriz}.  First we remove a section from $-17<M<-18$ in case (1) and then extend this region to $M=-19$ in case (2). Whilst there are slight differences in {\tctv}, compared with the results for the full datatset, toward the faint end, there is no overall systematic departure from completeness indicated.  Once again, by randomly replacing in the absolute magnitude slices a fraction of the removed objects both estimators show no change -- as can be seen in the bottom right panels.

In this and the previous section we have demonstrated the conditions under which our completeness test will remain robust.  In the following sections we now consider cases where on the other hand we expect the estimators to be very sensitive to incompleteness.

\subsection{Breaking separability}\label{sec:breaksep}
We can now explore in more detail the circumstances under which the assumption of separability breaks down.  This can help us characterise features observed in the completeness results which take us beyond the simple notion of finding the true apparent magnitude limit of a survey -- in essence observing and quantifying systematic departures from completeness. Figures~\ref{Fig:sys2} and \ref{Fig:sys2_mock1} demonstrate how basic forms of incompleteness, in {\it apparent} magnitude, will manifest themselves in {\tctv}.

\subsubsection{Unobserved objects: `under-completeness'}\label{sec:undecomp}
\begin{figure*}
	\begin{center}
	   \includegraphics[trim=0 0 0 3cm, width=0.49\textwidth]{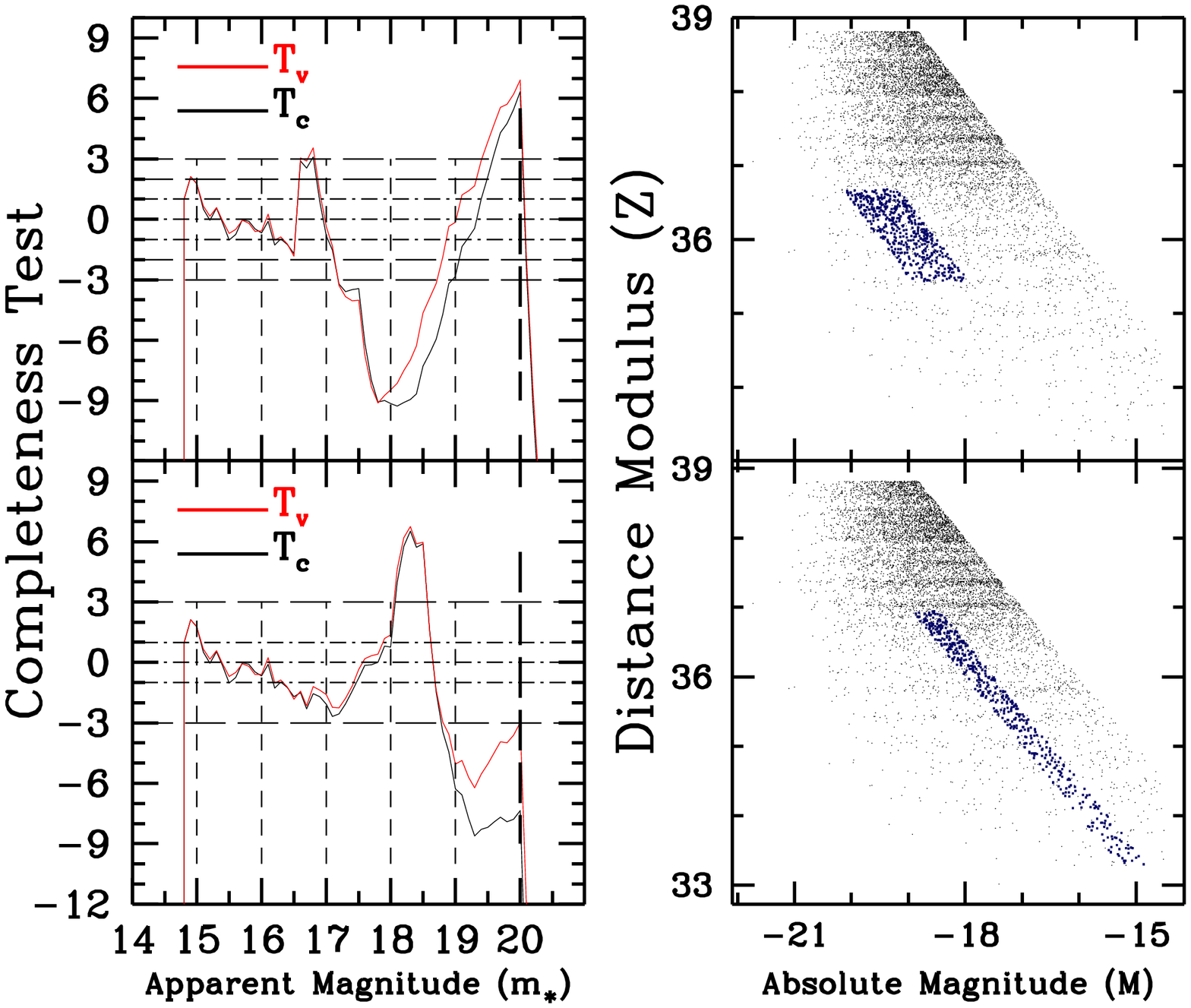} \hfill
	    \includegraphics[trim=0 0 0 3cm, width=0.49\textwidth]{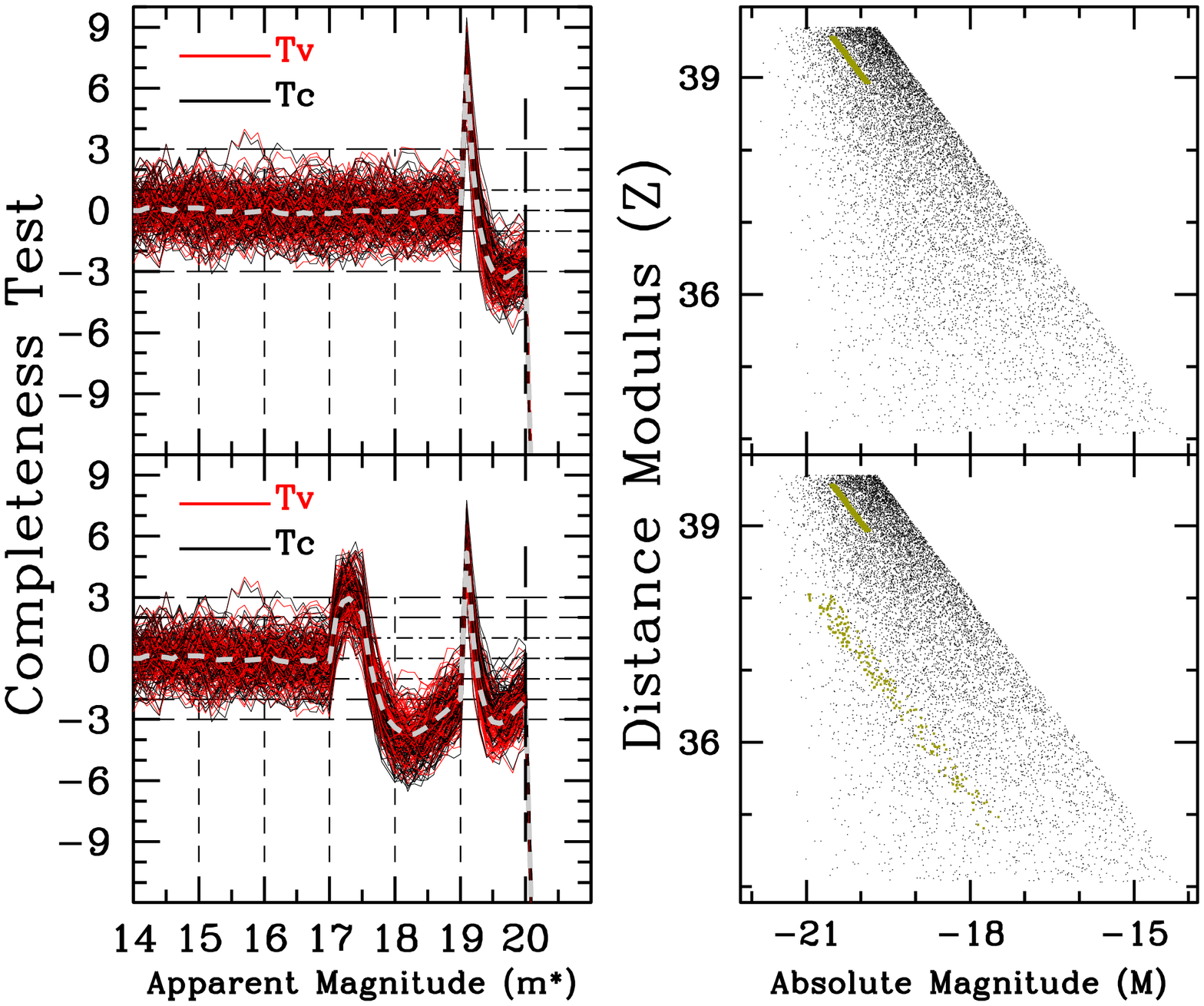}
	      \caption{\small  Adding artificial systematics.  The left-hand panel sets show the resulting {\tctv} curves when we add objects in a given apparent magnitude range to the MGC survey data. The top panels in this set have had 500 objects added randomly within the range range $16.55<m<17.5$ and $0.04<z<0.07$.  In the bottom panels of this set  the same number of galaxies were added in the range $18.0<m<18.5$ and $0.015<z<0.079$.  In the right-hand panel set we have taken our 100 Universal mock catalogues and added to each a similar form of systematic deviation. The top panels of this set show {\tctv} when 400 objects have been randomly added in the range $19.0<m<19.1$ and $0.13<z<0.17$. In the bottom panels we now add a further 200 galaxies between $17.0<m<17.5$ and $0.02<z<0.09$. In both cases the averaged {\tc} values are shown as a grey dashed line.}
	      \label{Fig:sys2_mock1}
\end{center}
  \end{figure*}

In the first scenario we remove all galaxies from a section of the {\mz} plane within a specified range of apparent magnitude $16.5<m<18.5$ and a distance modulus range $35.5<Z<37.5$ (Figures~\ref{Fig:sys2} top-left panel set).  From a parent set of 7847 galaxies, we have removed a total of  1048, leaving 6799 galaxies.  As we can see from the {\mz} distribution on the upper right-hand of  this panel set, these limits bound a parallelogram `slice' the removal of which thus introduces an artificial correlation between $M$ and $Z$.  Of course, this an unrealistic systematic effect, but it provides a useful, if exaggerated,  example to demonstrate the characteristic  behaviour of {\tctv} when galaxies have not been observed in a given apparent magnitude region.  We see that the removal of these galaxies produces consistent  behaviour in both {\tctv}: up to $m_*=16.5$ both estimators are consistent with sample completeness, as we would expect. However, as $m_*$ moves across the region of missing objects the estimators indicate significant incompleteness, dipping to a minimum of  $\sim-7\sigma$ between $16.5\lesssim m_*\lesssim 17.9$.   As {\mstar} now passes beyond the `incomplete' region, there is an immediate peak in the estimators at $\sim9\sigma$ between $18.5\lesssim m_* \lesssim 19.2$.  Such a feature appears to be a characteristic for this form of incompleteness, and is indeed to be expected as a direct result of the relation between the  separable regions in the construction of the  $\zeta$ and $\tau$ random variables.  In our rather extreme example we finally see the estimators drop below $-3\sigma$ before the `true' apparent magnitude limit of the survey indicated by the bold vertical dashed line.  We note at this point that in our {\it completeness I} paper  a similar trend was observed in our initial analysis of the 2dFGRS data.

In the bottom plots of the left-hand panel set of Figure~\ref{Fig:sys2} we now randomly return 538 galaxies, i.e. about half the removed galaxies, from the upper {\mz} distribution.  This results in a total subsample of 7309 objects.  Although we  now observe a much suppressed incomplete signal compared to the upper panels, the overall shape remains the same with the characteristic dip now observed at the $-\sim4~\sigma$ level followed by a peak at $\sim5.5~\sigma$. We also note, however, that the ability to constrain the faint apparent magnitude limit appears no longer to be impeded by the incompleteness at brighter magnitudes.

By using simple  Monte Carlo mock catalogues we can probe this effect in a more controlled environment. We construct a total of 100 mock catalogues, randomly drawing a total 10,000 objects from a Schechter LF and adopting LF parameters  as estimated in the MGC survey by \cite{Driver:2005} over a similar redshift range and out to a limiting magnitude of {\mlimf}=20.0~mag. Since we have already shown  our test statistics to be insensitive to clustering properties, for simplicity we have restricted our mock catalogues to have a uniform redshift distribution. In essence, our mock catalogues represent magnitudes drawn from a Universal LF where no systematics should be present.  The top plots in the right-hand panel set of Figures~\ref{Fig:sys2} show an example of an {\mz} distribution (right-hand plot) and the corresponding {\tctv} curves superimposed on each other for all 100 mock catalogues (left-hand plot).  The grey dashed and dotted lines shown in the left-hand plot are the respective averaged {\tctv} values which are both consistent with zero up to the faint limit.  Thus, it is clear from this plot that our mock catalogues are consistent with being complete up to the apparent magnitude limit of {\mlimf}=20.0~mag, as we would expect.

However, for each mock catalogue we now randomly  remove galaxies from two different ellipsoidal regions on the {\mz} plane. These regions are highlighted in blue and green.   From each mock this amounts to only a $\sim4\%$ fraction of the total number in the parent set. The impact of this removal is shown in the bottom plots of the right-hand panel set of Figures~\ref{Fig:sys2}.  We see that removing just a small fraction of objects results in a strong, systematic departure from completeness caused firstly by the green ellipsoidal region.  This feature is characterised by the dip in {\tctv} at {\mstar} values that correspond to the green region,  as indicated by the green vertical dashed lines on the completeness plot.  The following characteristic rise in the estimators  is suppressed slightly by the blue region as we observe a flattening of {\tctv} between $16.6\lesssim m_* \lesssim 17.7$ corresponding to this region of incompleteness and indicated on the figure by the blue vertical dashed lines.  There is then an overall peak in the both estimators at an average value of $\sim4.5 \sigma$ at {\mstar}$\sim18.2$, before both drop once again toward the faint limit.

The key points to take from this demonstration can be summarised as follows. If objects have not been observed,  for example,  in some apparent magnitude  range out to or within a given redshift range, our completeness estimators will show a consistent  and systematic characteristic shape that will typically lie outside the  expected statistical fluctuations.  This shape takes the form of a drop in both {\tctv} as {\mstar} moves across the incomplete region, followed by a distinct peak as {\mstar} moves toward fainter magnitudes and hence more objects. To test whether such a systematic effect is limited to a particular redshift range, one could of course split the {\mz} distribution into a series of redshift slices to probe where it is most prominent.

\subsubsection{Too many objects: `over-completeness'}
 We now consider the opposite scenario where one observes an excess of objects in an apparent magnitude bin relative to that expected from the overall population of galaxies.   Such a scenario, which we can refer to as {\it over\/} completeness, could be induced for example by magnification effects caused by gravitational lensing -- i.e. objects that would have ordinarily have been too faint to be observed  in a flux-limited survey are  lensed into the catalogue.

Again, for illustrative purposes, we randomly add objects in apparent magnitude to the survey sample, as shown in the left-hand panel set Figure~\ref{Fig:sys2_mock1}. In the top plots, 500 objects have been uniformly randomly added within $16.55<m<17.50$ and within a redshift range $0.04<z<0.07$.  This distribution is highlighted by the blue points in the {\mz} distribution. The resulting {\tctv} results are shown in the left hand panel.  The following systematic features are observed for both estimators.  As {\mstar} moves across the incomplete region a distinct peak is observed at the $\sim3\sigma$ level over the range $16.55\lesssim m_* \lesssim 16.8$. This is then followed by a systematic drop, reaching the $\sim-9\sigma$ level,  and finally  another peak at the $\sim6\sigma$ level just before the estimators drop sharply at the magnitude limit of the survey.
\begin{figure*}
	\begin{center}
	   \includegraphics[width=1.\textwidth]{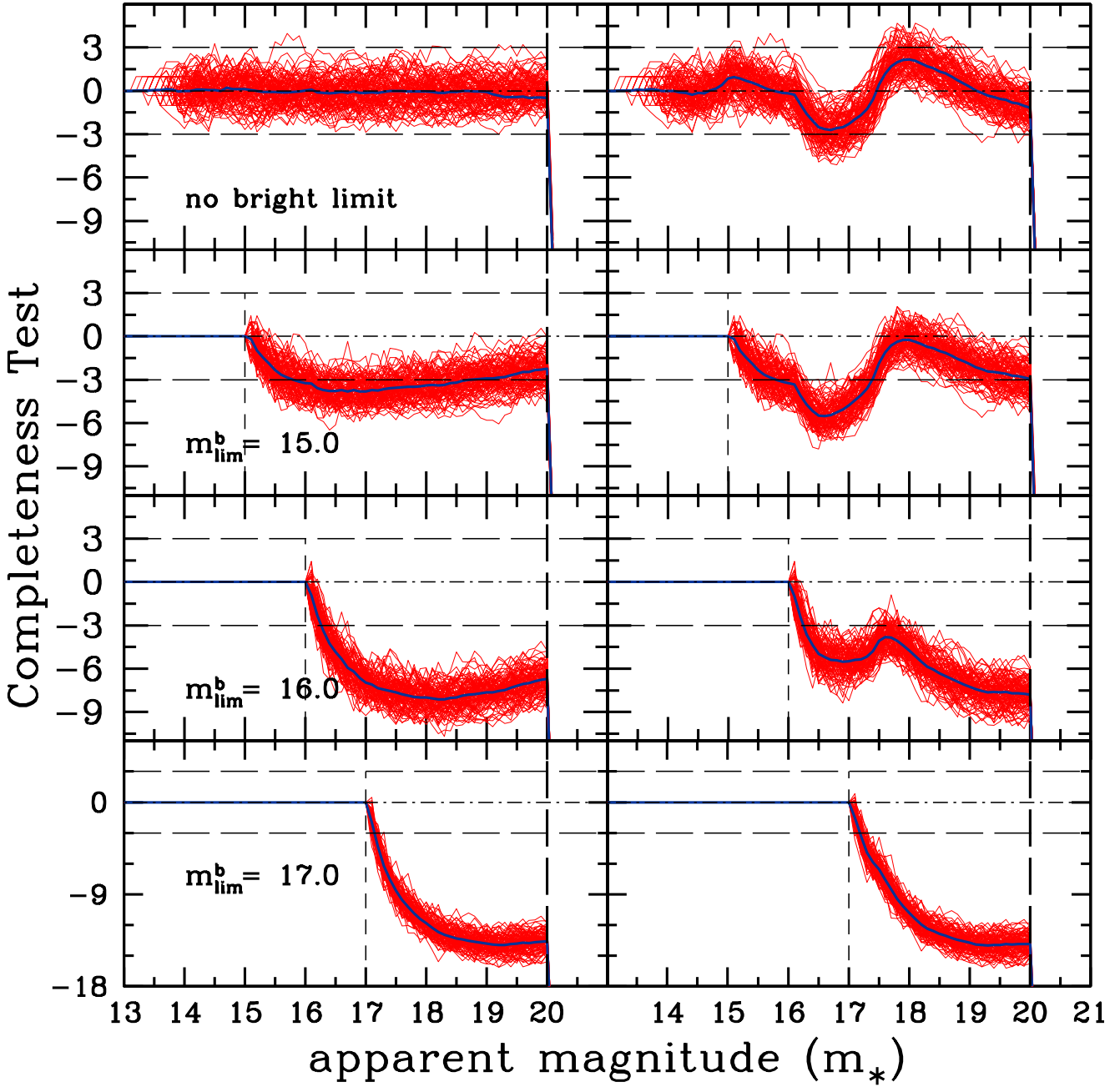}
	      \caption{\small Exploring the effects of imposing a bright apparent magnitude limit. The left- and right-hand panels show the same mock catalogues as used in Figure~\ref{Fig:sys2}:  the left hand panels show mock catalogues drawn from a Universal LF, and the right-hand panels show the same mock catalogues but with a systematic effect added to the apparent magnitude data. As in the previous illustrations, the red lines show {\tc} results for 100 mock catalogues and the dark blue line in each panel shows the average value of {\tc} from these 100 mock catalogues.  In each case we apply the Rauzy estimator to the data. From top to bottom we introduce successive cuts in apparent magnitude at increasingly fainter magnitudes, with the top panels showing the initial case of having no bright limit.  The remaining panels are respectively cut at {\mlimb}=15.0, 16.0 \&17.0 mag. The  key result in this figure is the demonstration that the overall features of the completeness test remain largely unaffected,  but do suffer 
a systematic downward shifting in {\tc} at apparent magnitudes brighter than the faint limit as we move from the case where there is no bright limit to a gradual well defined bright limit. }
	      \label{Fig:brightlim}
\end{center}
  \end{figure*}

In the bottom panels  of  Figure~\ref{Fig:sys2_mock1} we have now added the same number of objects as before but to a narrow magnitude range of  $18.0\lesssim m \lesssim 18.5$ and extending across a wider redshift range of $0.015<z<0.079$.  We observed a similar trend as with the previous example.  {\tctv} peaks  at the  $\sim6.5\sigma$ level  between  $18.0\lesssim m_* \lesssim 18.5$, which corresponds to the incomplete region.  We then observe a dip in the estimators beyond $m_*> 18.8$ at the  $\sim-6\sigma$ level for {\tv} and  $\sim-9\sigma$ level for {\tc}.  The estimators then show a slight rise toward the faint limit before dropping sharply at $m_*=20.0$.

As a check in a more controlled environment we once again manipulate our mock catalogues in a similar way by adding objects to them in a manner that breaks separability: see the right-hand panel set  in Figure~\ref{Fig:sys2_mock1}.  In the first example, we add 400 objects randomly in a narrow magnitude range $19.0<m<19.1$ and within $0.13<z<0.17$.  This case is highlighted in the {\mz} distribution in the top right panel in blue. We can see that both {\tctv} spike within the targeted magnitude interval, followed by a shallow dip at {\mstar}$\sim19.5$.  In the bottom panels we now extend our example by randomly adding another 200 objects in the intervals  $17.0<m<17.5$ and $0.02<z<0.09$. Once again we can see that both {\tctv} respond in a similar way to the addition of both systematic regions on the {\mz} plane.  We observe that the added systematic in this case marginally reduces the overall amplitude of the initial systematic spike at {\mstar}$\sim19.1$.

In summarising this section, it is clear to see that over densities in apparent  magnitude that break separability  have the effect of producing a positive spike in the completeness estimators at apparent magnitudes corresponding to the `over' complete region, followed by a dip at fainter magnitudes. This is, in essence, the opposite effect to that seen when galaxies were artificially removed from the sample to represent incompleteness.  Finally, one can expect to observe a second rise in the estimator, the amplitude of which depends where the incomplete region occurs.  

Note that in both examples of artificially induced systematic effects investigated above, when they are added to the real survey data in Figure~\ref{Fig:sys2_mock1}, the completeness estimators continue to show a sharp drop at the actual $m_{\lim}$ of the survey -- albeit in the second case of `over' completeness this occurs below the usual $-3\sigma$ limit.  The robustness of this effect was confirmed by our use of 100 mock catalogues.  In reality of course, survey samples are likely to be subject to varying degrees of incompleteness, manifesting themselves as features that may partially cancel out when incompleteness is probed across discrete magnitude bins. However, the illustrative examples in this section indicate that one could adopt an appropriate weighting scheme to correct for these localised incompleteness features, where the weight assigned to each apparent magnitude bin is directly related to the value of the completeness estimators in that bin. We will investigate this approach in detail in a future paper.

\subsection{Revisiting the bright limit case}

One of the key developments in  {\CI} was extending the completeness test to account for the presence of a bright apparent magnitude limit in galaxy surveys. This development was motivated in part by the completeness results obtained when analysing a galaxy sample from the 2dFGRS.   Through the use our mock catalogues from the previous sections we  now explore more deeply how the presence on an unmodelled bright limit affects the behaviour of the completeness estimators and how this behaviour relates to the results in  {\CII}, where we showed that choosing a {\dM} and {\dZ} that is too small can mask underlying incompleteness.

In the left-hand panels of Figure~\ref{Fig:brightlim} we have taken the mock catalogues drawn from a Universal Schechter LF and applied a succession of three increasingly faint bright limits in apparent magnitude at {\mlimb}=15.0, 16.0 and 17.0 mag.  For this demonstration we present only the results for {\tc} and  note that {\tv}  displayed the same behaviour. The red lines in the figure represent the results for each mock catalogue and the blue line in each panel shows their average.  In all cases {\tc} is calculated using the Rauzy estimator -- i.e. no bright apparent magnitude limit is modelled when computing the statistic, even when a bright limit is present in the data. The top-left panel shows the results when we do not impose any bright limit on the dataset and the mock catalogues are thus complete within and up to the faint apparent magnitude limit {\mlimf}=20~mag. As we begin to  impose a bright limit we observe a  systematic downward shift  in {\tc} to increasingly negative values.  Despite this behaviour, the sharp downward trend in {\tc} beyond the true faint apparent magnitude limit remains unaffected.  Moreover, the presence of an unmodelled bright limit does not induce on {\tc} any of the systematic characteristics shown in the previous sections.

To explore this behaviour further we then applied the same bright apparent magnitude cuts to the mock data samples but this time with a systematically under-sampled region included in each mock catalogue, in the manner described in the previous section. This was achieved by randomly removing the single ellipsoidal region corresponding shown in green on the right hand panel set of Figure~\ref{Fig:sys2}. Again in all cases {\tc} is calculated using the Rauzy estimator. The top-right panel shows the results of the completeness test when no additional bright magnitude limit is imposed.  The right hand panels below then show the results of imposing an increasingly deeper bright limit on the data.  As we cut deeper in magnitude to {\mlimb}=15.0 and 16.0 we can clearly see the same downward shift in {\tc} evident in the left hand panels. Perhaps more importantly, the systematic dip at {\mstar}$\sim17$ caused by the under-sampled region in the mock catalogues also remains.  In the bottom right panel, with a bright apparent magnitude limit at {\mlimb}=17.0, we see that {\tc} now exhibits the same trend as the in the corresponding left-hand panel. This is because the bright cut in apparent magnitude is in this case already fainter than the region where the systematic undersampling is introduced.

We conclude, therefore, that if a survey sample has an unmodelled bright apparent magnitude limit then our completeness test statistics will be systematically shifted below $-3\sigma$ over a wide range of trial faint apparent magnitude limits until {\mstar} moves beyond the true faint limit  of the survey -- at which point the test statistics will again drop very sharply.  Moreover, any localised systematic features in {\tc} or {\tv}  resulting from incomplete survey data will still be present. Thus, the characteristic incompleteness results which we obtained in {\CI} when we analysed the 2dFGRS data would appear to be the result of not only an unmodelled bright magnitude limit, but also the presence of additional systematic effects in the survey, of the form illustrated in \S~\ref{sec:undecomp}.

\subsection{A probe of luminosity evolution}
Source evolution in survey data can be thought of as  another form of incompleteness.  There are a number of  methods available to constrain the statistical properties of  evolution for a population of galaxies.  Probably the most common approach is to parameterise evolution with a redshift $z$ dependent model  where either pure luminosity evolution (PLE)  [e.g. $L_*(z)=L(0)(1+z)^k$] or pure number density evolution  [e.g. $\phi_*(z)=\phi(0)(1+z)^\gamma$], or a combination of both, is inferred from the estimated luminosity functions as a function of redshift.  In the PLE case it is generally assumed that galaxies were brighter in the past where $L*$ is the characteristic luminosity of the LF  and $k$ is a galaxy type dependent evolution parameter.  Similarly, with number evolution $\gamma$ is an evolution parameter which assumes galaxies were more numerous in the past,  where $\phi_*$ is the normalisation of the LF.   It is common practice to constrain these models using a maximum likelihood estimation (MLE) technique, involving an assumed parametric form for the  LF  \citep[see e.g.][]{Saunders:1990,Heyl:1997,Springel:1998,Croom:2004MNRAS.349.1397C,Wall:2008MNRAS.383..435W}. In both cases it is assumed when carrying out the MLE that the underlying evolutionary model is the correct one to describe the entire population of galaxies in the sample under test.

In this section we briefly show how our completeness estimators can be used as a probe of PLE without requiring any knowledge of the parametric shape of the LF.  In a follow up paper we will demonstrate how we can adopt this approach to constrain the parameter(s) of a PLE model, although again without requiring any parametric model for the LF.  

For this initial study we draw upon the work of \cite{Croom:2004MNRAS.349.1397C} (hereafter C04) who constrained evolutionary models for high redshift quasi-stellar objects (QSO) over a broad redshift range.  As is common with QSO studies they adopt a two power-law LF of the form,
\begin{equation}\label{eq:LFcroom}
\Phi(M,z)=\frac{\Phi^*}{10^{0.4(\alpha+1)(M-M^*)}+10^{0.4(\beta+1)(M-M^*)}},
\end{equation}
where $M^*$ is the characteristic absolute magnitude, $\Phi^*$ is the normalisation, and \textcolor{\red}{$\alpha$ and }$\beta$ determine the slopes of the respective power laws \citep[see also][]{Boyle:1988MNRAS.235..935B}. In C04 they then characterise evolution as a second order polynomial expressed in magnitudes as,
\begin{align}\label{eq:Ecroom}
M^*(z)&=M^*(0)- E(z)\\ \nonumber
&=M^*(0)-2.5(k_1z+k_2z^2),
\end{align}
where $k_1$ and $k_2$ are the evolution parameters which are analogous to the $\beta$ evolution parameter described earlier in this section. In our study we use the C04 survey data as  a guide to create mock  QSO catalogues which have sufficient depth in redshift to probe evolution. We provide only the details of these data which are pertinent to this study. For further details please refer to C04 and references therein. 
In this scenario we assume that the $k$-correction is known. Such a correction is required since galaxies are observed at  different redshifts making use of single band filters thus sampling a fraction of the total spectrum.  To compare the measurements  at different redshifts one will need to convert the observation to the  rest frame of the object and therefore to correct for the finite size of the filter(s). In practice $k$-corrections are approximated using two-dimensional polynomials as function of a redshift and observed colour. Effectively, $k$-correction and PLE are degenerate
and will impact in the overall {\tc} behaviour in a similar manner.
Another subtle issue concerning the $k$-correction is the way one defines the areas $S_1$ and $S_2$ and the magnitude limits themselves: if the galaxies were first selected and then  $k$-corrected such limits would become  ``fuzzy" since different galaxy types and intrinsic colours would require different corrections at a given redshift. However, the usual procedure is to apply a magnitude cut after $k$-correction. Thus our implementation of {\tc}  adopts a ``top-hat"  approach to partition the data-sets rather than using ``fuzzy" magnitude limits.

The parent data-set comprised of 23,338 QSOs within the apparent magnitude range $18.25 <b_J<20.85$.  The  bright-end of the luminosity function was essentially extended with the introduction of a further 1564 QSOs in the range  $16.0 <b_J<18.25$ from the  6dF QSO redshift survey data-set.  The redshifts were selected  in the redshift range $0.4 < z <2.1$. In C04 the authors impose a further cut removing all sources $M > -22.5$.  With this subset they then apply a maximum likelihood technique to simultaneously constraint the LF and E(z) parameters of Equations~\ref{eq:LFcroom} and \ref{eq:Ecroom}. The results of this analysis are reproduced in the top row of Table~\ref{table1}.
\begin{table}
\begin{center}
\caption{\small Luminosity and evolution function parameters used to generate Monte Carlo mock magnitudes over the redshift range $0.4 < z < 2.1$ }
\label{table1}
\tabcolsep 3pt

\begin{tabular}{lcccccccc} \\ \hline \hline
\\
{Details}  	& {M$_{\lim}$ }   & {$M^*(0)$ }    & {$\alpha$}  & {$\beta$}   	&{$k_1$}   & {$k_2$} &   $m^f_{\lim}$
\\
\\
\hline
\\
C04               		&-22.5 & -21.61  &  -3.31  &   -1.09  	& 1.39 & -0.29   &  20.85 & \\
Universal LF              &--        &  -22.8  &  -3.31    &   -1.09     &  --     &    --      &   20.85 &\\
Evolved LF          	&--        & -21.61 &  -3.31   &   -1.09    &   1.39 & -0.29  &   20.85 &\\
\\
\hline \hline
\end{tabular}
\end{center}
\end{table}
\subsubsection{The no evolution case}

 \begin{figure}
	\begin{center}
	  \includegraphics[width=0.48\textwidth]{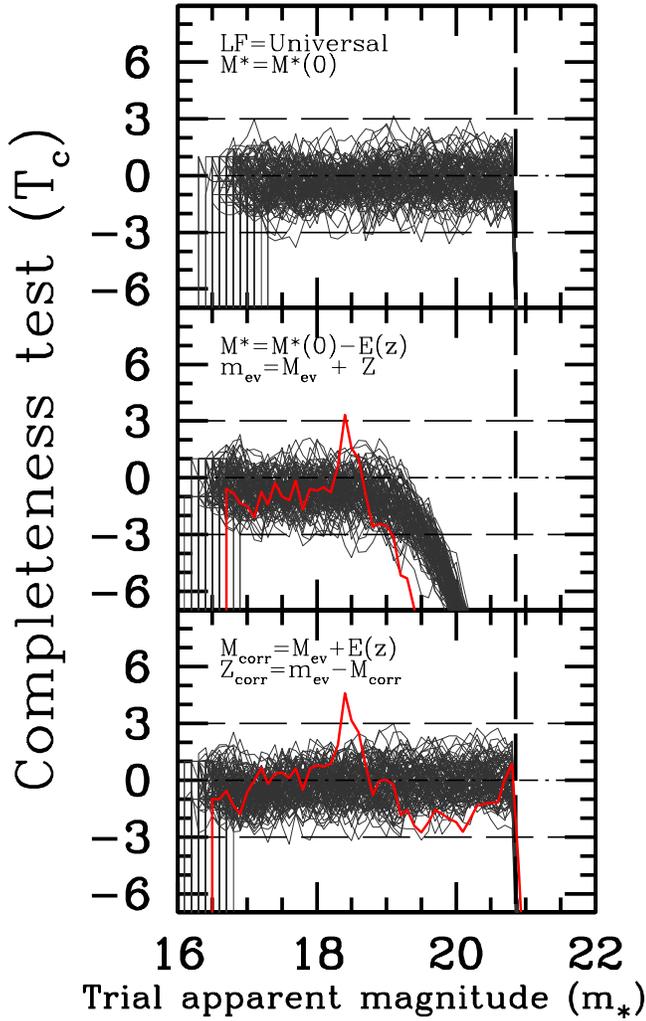}
	      \caption{\small Probing  pure luminosity evolution (PLE). {\it Top panel} - Each of the 100 mock catalogues (grey lines) were drawn from a Universal (no evolution)  two-power law luminosity function as described in the text.  As we would expect from a mock catalogue without an evolving $M^*$ term in the luminosity function, all the mock catalogues show consistent completeness up to the faint apparent magnitude limit of $m=20.85$~mag. {\it Middle panel} - we now generate absolute magnitudes for our mock catalogues from an evolving luminosity function, where $M^*$ is allowed to vary according to a redshift dependent PLE model taken from  C04. This model effectively breaks the separability between $M$ and $Z$ and the resulting {\tc} curves for each mock catalogue show a characteristic drop below $-3\sigma$ at approximately $m_*\sim 19.6$. The red line shows the completeness statistic for the actual C04 QSO data prior to applying any evolution correction.  We can see the same characteristic drop-off, indicating the presence of evolution, for the real data. {\it Bottom panel} - Finally, we now correct our evolved mock data with the evolution function that was originally used to generate the magnitudes from the evolving luminosity function.  The grey lines, for the mock catalogues, now show results consistent with a complete sample. Similarly, correcting the C04 QSO data with their constrained evolution parameter values we can see their data is now consistent with completeness up to their published limiting magnitude of $m=20.08$~mag.}
	      \label{Fig_sys3}
\end{center}
  \end{figure}

In the first step of our analysis we create a set of control samples by generating mock catalogues drawn from a Universal LF. Throughout we adopt the same cosmology as in C04 such that $\Omega_m=0.3$, $\Omega_\Lambda=0.7$ and $H_0=70$~kms$^{-1}$ Mpc$^{-1}$. Essentially this means that we do not add the evolution term of Equation~\ref{eq:Ecroom} and thus the assumption of separability between $\Phi(M)$ and $\rho(Z)$ should remain valid for our control samples. We note that the constrained LF parameters from C04 are derived from observed magnitudes that are subject to evolution.  As Table~\ref{table1} shows, we adopt the same LF parameters as C04 with the exception of setting a slightly brighter $M^*(0)$ value of $-22.8$~mag to produce a more sensible magnitude distribution.

For simplicity, we generate a uniform random sample of redshifts in the range $0.4 < z < 2.1$  throughout this study. In future work we will  introduce clustering effects; however it should be noted that, by design, our completeness estimators are insensitive to clustering.  Furthermore we do not impose any cuts in absolute magnitude.   Thus, for an object at redshift $z_i$ we randomly sample  an absolute magnitude from the cumulative distribution function (CDF) of the adopted LF Equation~\ref{eq:LFcroom} and compute the following,
\begin{align}
Z_i &= 5\log_{10}(d_L)+25\\
m^{\rm uni}_{\rm samp}(z_i)&= M^{\rm uni}_{\rm samp}(z_i) +Z_i
\end{align}
where $m^{\rm uni}_{\rm samp}$ denotes the  sampled apparent magnitude from a universal LF, $Z_i $ is the distance modulus and $d_L$ is the luminosity distance.  The superscript {\it uni} refers to sampling from a Universal LF.  Final selection of a galaxy must meet the following condition,
\begin{equation}
m^{\rm uni}_{\rm samp}(z_i) < m_{\lim}^{\rm f}(\mbox{survey}),
\end{equation}
where  $m_{\lim}^{\rm f}$ is defined by the C04 QSO survey limit.

From this starting point we generate 100 realisations each comprising a total of 18662 objects. 
With no redshift dependency in the absolute magnitudes generated for our mock surveys, they should of course be magnitude complete. To verify this we compute the {\tc} statistic for each mock survey. The results are shown in the top  panel of Figure~\ref{Fig_sys3},  where we have superimposed each of the 100 {\tc}  curves onto this plot, shown in grey. As expected, our results show no indication of incompleteness or other systematic effects, with the {\tc} statistic for each mock catalogue dropping sharply beyond the apparent magnitude limit of 20.85~mag.
\subsubsection{The evolution case}
We now consider the case of an evolving LF. More explicitly, Equation~\ref{eq:LFcroom} now becomes,
\begin{eqnarray}\label{eq:LFcroom_ev}
{\Phi(M,z)} =
\frac{\Phi^*}{10^{0.4(\alpha+1)[M-M^*(z)]}+10^{0.4(\beta+1)[M-M^*(z)]}}.\quad
\end{eqnarray}
Thus, for each object located at $z_i$ a unique CDF for $M$ is generated from which a mock absolute magnitude is sampled.  Note that we adopt the same redshift distribution as in the no-evolution case.

Table~\ref{table1} shows the parameters adopted for the LF and the evolution term.  For clarity, in this scenario for each `evolved' absolute magnitude sampled, $M^{\rm ev}_{\rm samp}$,  we compute,
\begin{align}
Z_i &= 5\log_{10}(d_L)+25\\
m^{\rm ev}_{\rm samp}(z_i)&= M^{\rm ev}_{\rm samp}(z_i) +Z_i,
\end{align}
where the superscript {\it ev} reminds us that the magnitudes are drawn from an evolving LF.

When we now compute {\tc} for our modified set of mock catalogues we do not expect a result consistent with completeness.  The middle panel of Figure~\ref{Fig_sys3} confirms this, showing in grey the resulting {\tc} curves for this suite of mock catalogues. In clear contrast to the non-evolved case of the previous section we now observe a steady and systematic decline of {\tc} for each mock catalogue, with the average of the grey lines crossing the $-3\sigma$ threshold at  $m_*\approx19.6$~mag, i.e. 1.25 mag. brighter than the true faint limit of the datasets.  We note that there are no other signs of systematic departure from completeness observed between $16.0 \lesssim m_* \lesssim 19.6$.

The red line on the middle panel shows the {\tc} statistic computed for the C04 observed QSO sample (uncorrected for evolution). It is reassuring to note that we observe the same characteristic drop in {\tc} as with the mock catalogues, albeit at a slightly brighter $m_*$ value of 19.30~mag. We also observe a small spike in the C04 {\tc} curve at {\mstar}$\sim18.4$ which may be due to the transition to the combined 6dF data.
\begin{figure*}
	\begin{center}
	 \includegraphics[width=0.33\textwidth]{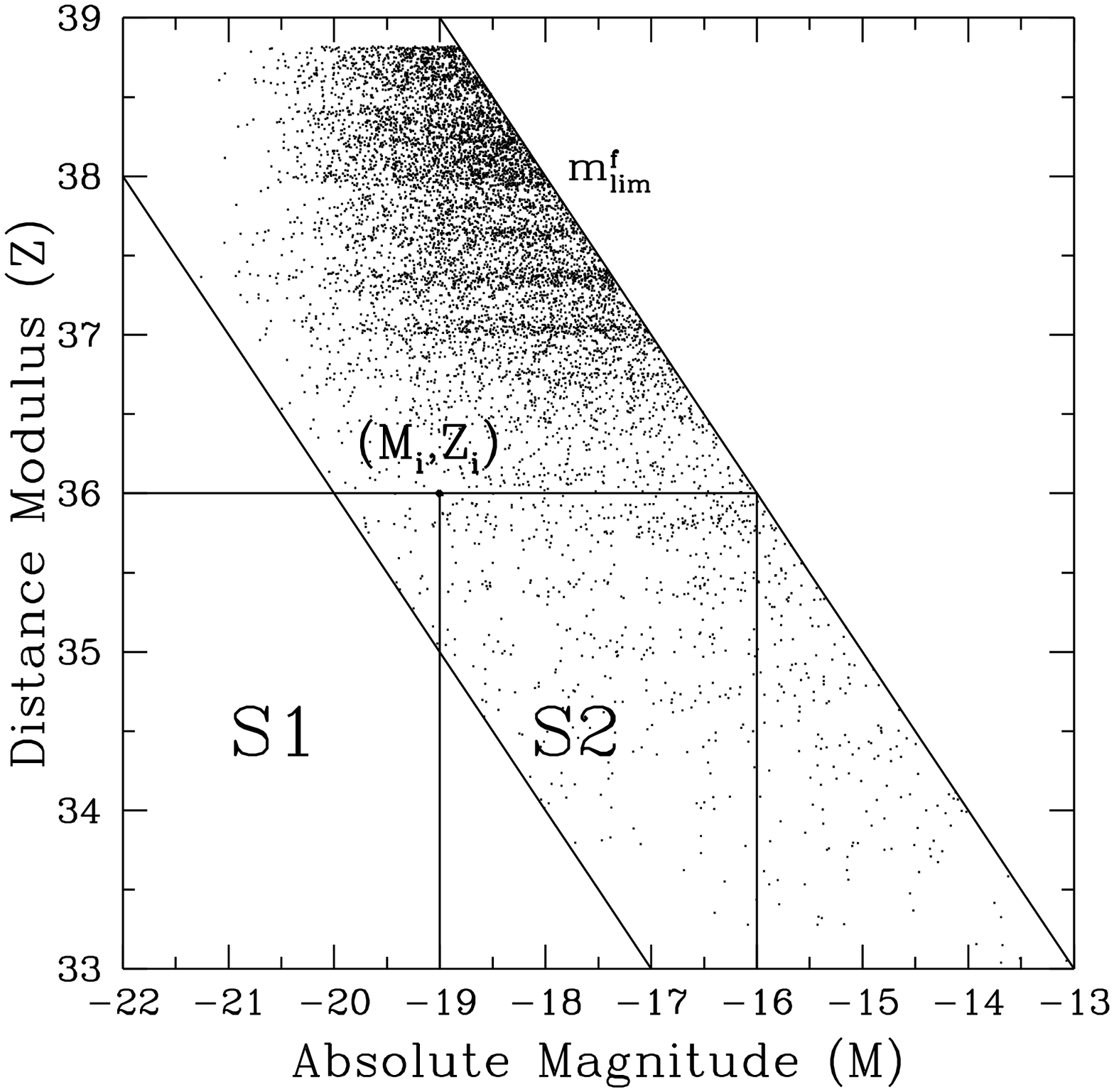}
	  \includegraphics[width=0.32\textwidth]{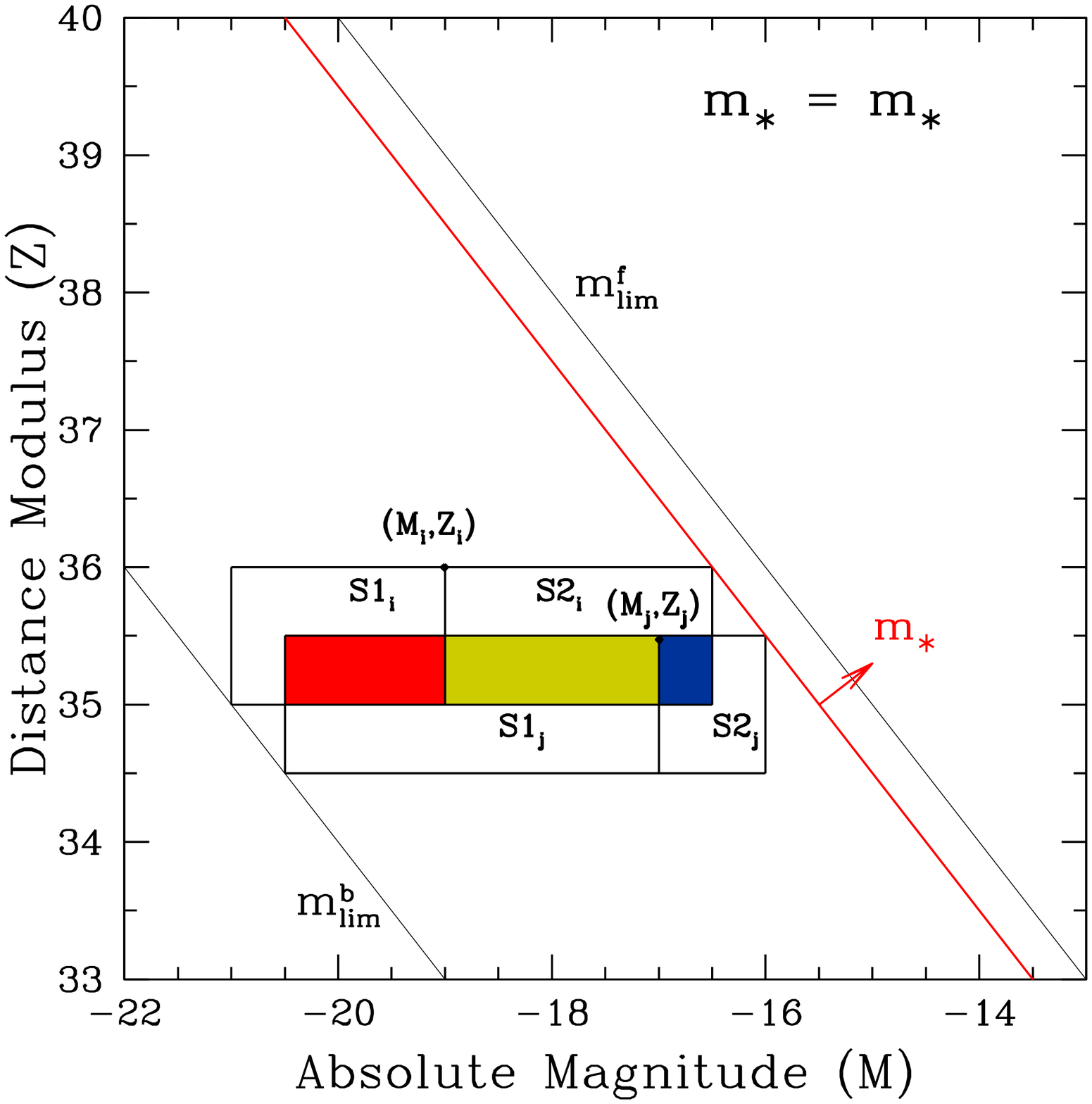}
	  	  \includegraphics[ width=0.32\textwidth]{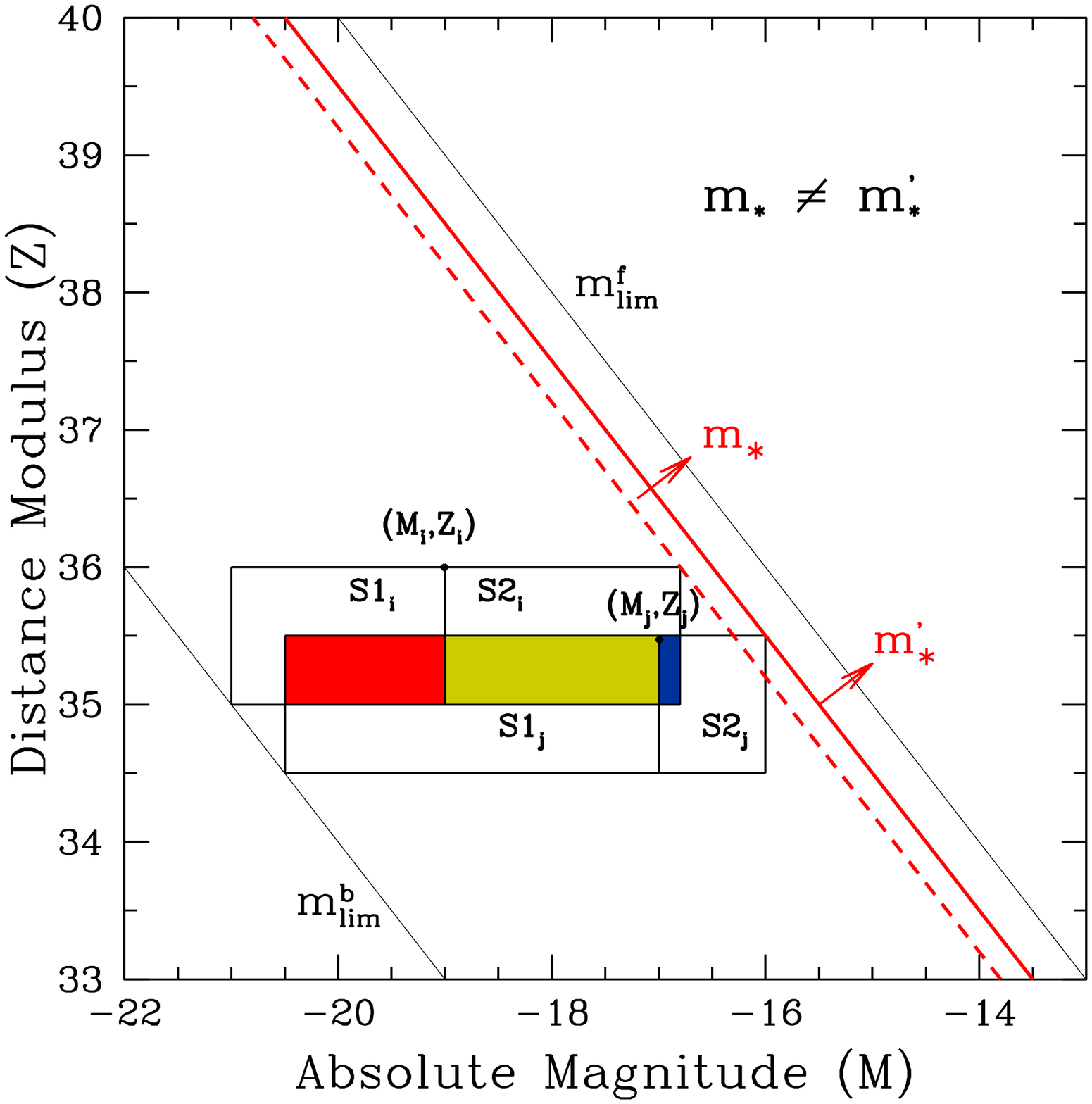}

	      \caption{\small   Illustrating the manner in which we account for the correlations between certain $\zeta$ calculations. Left panel:  an illustrated {\it complete} data-set with well defined cuts in apparent magnitude at both the faint and bright end.  Applying the traditional {\it Rauzy} (R01) approach, one can see that the $S1$ and $S2$ regions are no longer separable due to the presence of the bright limit.   Middle panel: in this example we have shown two arbitrary regions that would be used to estimate $\zeta_i$ and $\zeta_j$ for respective galaxies located at ($M_i,Z_i$) and ($M_j,Z_j$) for a given {\mstar}.  As such both galaxies form the respective regions [$S1_i, S2_i$] and [$S1_j, S2_j$].   We observe by the coloured regions that there is a correlation introduced into the overall $\zeta$ calculation where these regions overlap.   That is, $\zeta_i$ and $\zeta_j$ are not independent. In this particular case we have the following scenarios where: [$S1_i\cup S1_j$] (red region),  [$S2_i\cup S1_j$] (green region) and  [$S2_i\cup S2_j$] (blue region). Right panel: we now consider correlations when {\mstar}$\ne m_*'$. That is, we compare the ($M_i,Z_i$) region for {\mstar} and the subsequent  ($M_j,Z_j$) region for a fainter {\mstar} value denoted by  $m_*'$.}
	      \label{Fig:zetacorr}
\end{center}
  \end{figure*}

\subsubsection{Correcting for evolution}
To conclude this section we now consider the effect of correcting the evolved mock samples with the evolution model used to generate them. In this way we now compute,
\begin{align}
M_i^{\rm corr} &= M^{\rm ev}_{\rm samp}(z_i) + E(z)\\
Z_i^{\rm corr} &=m^{\rm ev}_{\rm samp}(z_i) - M^{\rm corr},
\end{align}
where the superscript {\it corr} represents the data corrected for evolution. In the absence of any other systematic effects, the correct evolutionary model applied as above should now render the {\mz} distribution separable and should thus produce a test statistic that is once again consistent with completeness.  If we now look at the bottom  panel of Figure~\ref{Fig_sys3} we can see that this is indeed the case.  By correcting the evolved {\mz} distribution with the appropriate evolutionary model, the {\tc} curves are now consistent with the results obtained for the Universal case, in the top panels.

Finally, we correct the actual C04 data in the same way.  The resulting {\tc} statistic is shown in red in the bottom panel.  We again see that the corrected C04 data are consistent with completeness up to the published apparent magnitude limit.

In a follow-up paper we shall explore in more detail how we can use our completeness estimators to constrain PLE models by either assuming a parametric model of evolution or as a free-form technique.  In both cases the important feature of our method is that no parametric form for the LF, or indeed for the spatial distribution of the sources, is required.

\section{{\tctv} error propagation}\label{sec:errorprop}

Whilst the optimisation in {\CII} is useful for maximising the signal to noise ratio during the sampling process, a useful further step would be to understand the  error on both {\tctv} either for a given {\sn} level or when applying the R01 or {\CII} approaches.  

Let us first re-define $\zeta_i$ for survey object $i$ as function of $m_*$: 
\begin{equation}
\zeta_i(m_*) \equiv
\begin{cases}
0, & \mbox{for objects which one cannot} \\
&\mbox{compute $r_i(m_*)$ and $n_i(m_*)$,}\\ \\
\frac{r_i(m_*)}{n_i(m_*)+1}
, & \mbox{otherwise,}
\end{cases}
\end{equation}
recalling that $r_i$ is the number of objects in $S_1$ and $n_i$ is the number of objects in $S_1$  $\cup$ $S_2$.

Thus from the expression for $T_c(m_*)$, Equation~\ref{equ:tc},  
the error propagation can then be shown to be,

\begin{eqnarray}\label{Eq:fullerr}
\Delta T_c(m_*)&=& \sum _{i=1}^{N_{\rm gal}}  
 \left \{ \frac{ \Delta \zeta_i(m_*)}{ \Xi\left( m_*  \right) }  + 
\frac{T_c(m_*)\Delta n_i(m_*)}{ 2\,\Xi\left( m_*  \right)}  \frac{\partial }{\partial  n_i} {\rm Var} \left[\zeta_i(m_*) \right] \right \}  \nonumber \\
&=& \sum _{i=1}^{N_{\rm gal}}  \frac{1}{\Xi\left( m_*  \right)}
\left\{ \Delta \zeta_i(m_*) + \Psi_i (m_*)\Delta n_i(m_*)  \right\}
\end{eqnarray}

where 
$$ 
\Xi(m_*) \equiv \left \{ \sum_{i=1}^{N_{\rm gal}}  {\rm Var }\left[ \zeta_i (m_*) \right] \right\}  ^{1/2},
$$
and 
$$\Psi_i(m_*)\equiv \frac{T_c  (m_*)}{2}  \left \{\frac{\partial }{\partial  n_i} {\rm Var} [\zeta_i (m_*)] \right \} $$ 
Thus
the covariance  between two  $T_c$'s corresponding to two  magnitude limits, 
$m_*$ and $m_*'$, is given by,
\newline
\noindent$\left < \Delta T_c(m_*) \Delta T_c(m_*') \right >=$
\begin{align}\label{cov_matrix}
& \sum_{i,j}^{N_{\rm gal}} \frac{1}{\Xi(m_*) } \frac{1}{\Xi(m_*') }
\left \{ \left < \Delta \zeta_i (m_*) \Delta \zeta_j (m_*') \right >  + \right . \nonumber \\
& 
+\left< \Delta  n_i(m_*) \Delta n_j(m_*') \right > \times \Psi_i(m_*) \Psi_j(m_*')
- \nonumber \\
 &  -\left< \Delta n_i(m_*) \Delta \zeta _j(m_*') \right> \times \Psi_i(m_*) - \nonumber \\
 &  - \left . \left< \Delta n_i(m_*') \Delta \zeta _j(m_*) \right> \times \Psi_i(m_*')  
 \right\} .
\end{align}

The last three terms of the above expression contain factors of the random variable $\zeta_j -1/2$ [via $\Psi_i(m_*')$]. 
In the case that $m_*\ne m_*'$, these will lead to very small contributions to the overall covariance, since 
$\zeta_j (m_*)-1/2$ follows a uniform distribution centred around zero. However, if $\zeta_j$ is not uniformly distributed on the interval $[0,1]$ then the terms containing $\zeta_j -1/2$ will not necessarily vanish. This scenario will occur, for example, with the breaking of separability discussed in section~\ref{sec:character}.  Thus for a {\it complete\/} data-set and  for $m_*\ne m_*'$ (where $m_*$ and  $m_*'$ are both 
brighter than $m^{\mbox{\scriptsize{f}}}_{\mbox{\scriptsize{lim.}}}$, the true faint limit of the survey) the covariance matrix acquires a rather simple form:

\begin{align}\label{cov_matrix_simple}
\left < \Delta T_c(m_*) \Delta T_c(m_*') \right >= 
 12\sum_{i,j}^{N_{\rm gal}}  \frac{\left < \Delta \zeta_i (m_*) \Delta \zeta_j (m_*') \right >}{\left[ N_{\rm gal}(m_*) N_{\rm gal}(m_*') \right]^{1/2}}.\nonumber \\
\end{align}

Here we are also assuming that $n_i$ is large enough and consequently
$\Xi (m_*) \rightarrow \left[{N_{\rm gal}(m_*)}/12\right]^{1/2}$, where $N_{\rm gal}(m_*) \ne N_{\rm gal}$ since only objects brighter than 
$m_*$ are used to compute $T_c(m_*)$. The contributions from the individual objects $j$ and $i$ are shown schematically in  the right panel of Figure~\ref{Fig:zetacorr}. From Equation~\ref{Eq:zetasigtonoise} we can see that the computation of 
errors in the random variable $\zeta_i$ reflects the Poisson 
fluctuations of counting objects within areas $S_1$ and $S_1$ $\cup$ $S_2$.  Namely, the rank $r_i$ and $n_i$, respectively.

As for the case where $m_*= m_*'$ (see the middle panel of Figure~\ref{Fig:zetacorr}), the contribution of the term containing $\left< \Delta  n_i(m_*) \Delta n_j(m_*) \right >$ will be  restricted to $i=j$.  The last two terms of Equation~\ref{cov_matrix_simple} will still vanish for a large enough number of objects in the catalogue insofar as  the catalogue is complete and $m_*$ is brighter than the survey faint limit. The contributions from the individual objects $j$ and $i$ are shown schematically in the middle panel of Figure~\ref{Fig:zetacorr}.
\section{discussion \& future work}\label{sec:conclusions}
In the first part of this article we examined more closely the conditions under which systematic effects in survey data will manifest themselves in our completeness statistics.  To demonstrate this effectively, we firstly probed the conditions under which our completeness estimators remained robust to changes in the data.  Using actual survey data we uniformly randomly removed objects from the catalogue via a bootstrapping approach.  In a second test, we removed slices of data in either $Z$ or $M$.  This too resulted in no significant change in  both {\tctv}. Therefore, in both cases where separability between $M$ and $Z$ was retained, {\tctv} were shown to be robust.

Systematic effects in $m$ were then explored for  three cases which can be summarised as follows:
\begin{enumerate}
\item {\it Under-completeness}:  This may arise when objects have not been observed over a particular redshift range and within apparent magnitude bins.  Through the use of real and mock data, we demonstrated that if this effect is present at even the level of a few percent, a characteristic signature will be observed in {\tctv}, manifest as a significant drop in both estimators for trial apparent magnitude limits that lie within the incomplete region(s), followed by a steep rise in {\tctv} for fainter {\mstar} values.\\
\item {\it Over-completeness}: Here there may be more objects observed in a particular apparent magnitude range relative to the underlying distribution. In this case {\tctv} are observed to display the opposite characteristics to those of case (i). Thus, as {\mstar} moves across the incomplete region on the {\mz} plane, a distinct peak in both {\tctv} occurs followed by a distinct dip as {\mstar} moves back into the `complete' region.\\
\item {\it Evolution}: For the particular case of luminosity evolution studied here, we demonstrated that mock magnitudes which are  drawn from an evolving LF will show a characteristic and sustained drop in {\tc} below $-3 \sigma$  at an {\mstar} value significantly brighter than the actual apparent magnitude limit of the sample.  Applying the correct evolutionary model to both $M$ and $Z$, corrects this form of incompleteness. This characteristic behaviour for {\tc} was confirmed with the use of real QSO survey data from C04 which is known to display strong evolutionary properties. 
\end{enumerate}
We note, however, that in cases (i) and (ii) we used a highly artificial scheme to introduce systematics to the data. In reality, one might expect some combination of these incompleteness effects to be present in the data, which may therefore partially cancel out. Nevertheless one might adopt, for example, an iterative weighting scheme to correct for any residual signatures in {\tc} or {\tv}.  This approach will be investigated in a later paper in the context of estimating the galaxy luminosity function.

In the first part of this paper we also revisited the impact on {\tc} or {\tv} of surveys that have an unmodelled secondary bright apparent magnitude limit.   The purpose of this was also to revisit our initial completeness results for the 2dF survey which led to the extension of {\tc} for the bright limit case. We showed that imposing a successively fainter, but unmodelled, bright magnitude limit resulted in a systematic downward `shifting' in the value of the completeness statistics.  As the bright limit became fainter, the systematic shifting became more negative. We also showed that for the mock catalogues with a small systematic perturbation added to the apparent magnitude distribution, the systematic dip in {\tc} corresponding to the affected region also remained and was, if anything, more significant.  Thus, reconsidering our 2dF results from {\it Completness I}, we surmise that the systematic dip observed in the {\tctv} statistics was more likely the result of some other underlying systematic effect not related to the bright limit of the survey, but which was then subsequently masked by our choice of narrow width in $\delta Z$ and $\delta M$,  as explored in {\CII}.

In future work we will also explore how the properties of our random variables $\zeta$  and $\tau$ can be exploited to constrain luminosity evolution in galaxy surveys. In essence, $\zeta$ and $\tau$ are powerful probes for identifying residual correlations in $M$ and $Z$ due to evolution. Thus, to constrain evolutionary models, we can extend our methodology to include e.g.  the  Kullback-Leibler divergence \citep{Kullback:1951} relative entropy method, which measures the difference between two probability distributions $p$ and $q$, where $p$ represents the observed distribution of the data ($\zeta$ or $\tau$ in our case) and $q$ represents our theoretical model for that distribution (i.e. after application of a specific correction for luminosity evolution).

In the final part of this paper we presented a full error propagation and covariance analysis for the {\tctv} statistics, further developing the ``adaptive smoothing" procedure introduced in {\CII}.

By addressing all these issues we believe that we have laid  a very comprehensive foundation for testing magnitude completeness limits that will become crucial for the next generation of redshift surveys.

\section*{Acknowledgements}
RJ would (once again) like to thank David Valls-Gabaud and also Mat Smith  for their  insightful comments and fruitful discussions, and also acknowledges support from the National Research Foundation (South Africa) and the South African Square Kilometre Array. MAH would like to thank Tom Loredo and Woncheol Jang for useful discussions, and also acknowledges the Aspen Center for Physics where some of this research was carried out.

We would also like to thank the anonymous referee for his/her helpful comments that helped bring greater clarity to this work.

The Millennium Galaxy Catalogue consists of imaging data from the Isaac Newton Telescope and spectroscopic data from the Anglo Australian Telescope, the ANU 2.3m, the ESO New Technology Telescope, the Telescopio Nazionale Galileo and the Gemini North Telescope. The survey has been supported through grants from the Particle Physics and Astronomy Research Council (UK) and the Australian Research Council (AUS). The data and data products are publicly available from http://www.eso.org/~jliske/mgc/ or on request from J. Liske or S.P. Driver.
\setcounter{equation}{0}
\renewcommand{\theequation}{A-\arabic{equation}}
\setcounter{section}{0}
\renewcommand{\thesection}{A-\arabic{section}}
\setcounter{figure}{0}
\renewcommand{\thefigure}{A-\arabic{figure}}
\bibliography{bibliography_CIII}
\bibliographystyle{astron}
\section{Appendix: Creating Monte Carlo survey simulations}\label{app:mocks}
\begin{figure*}
     		\includegraphics[width=0.48\textwidth]{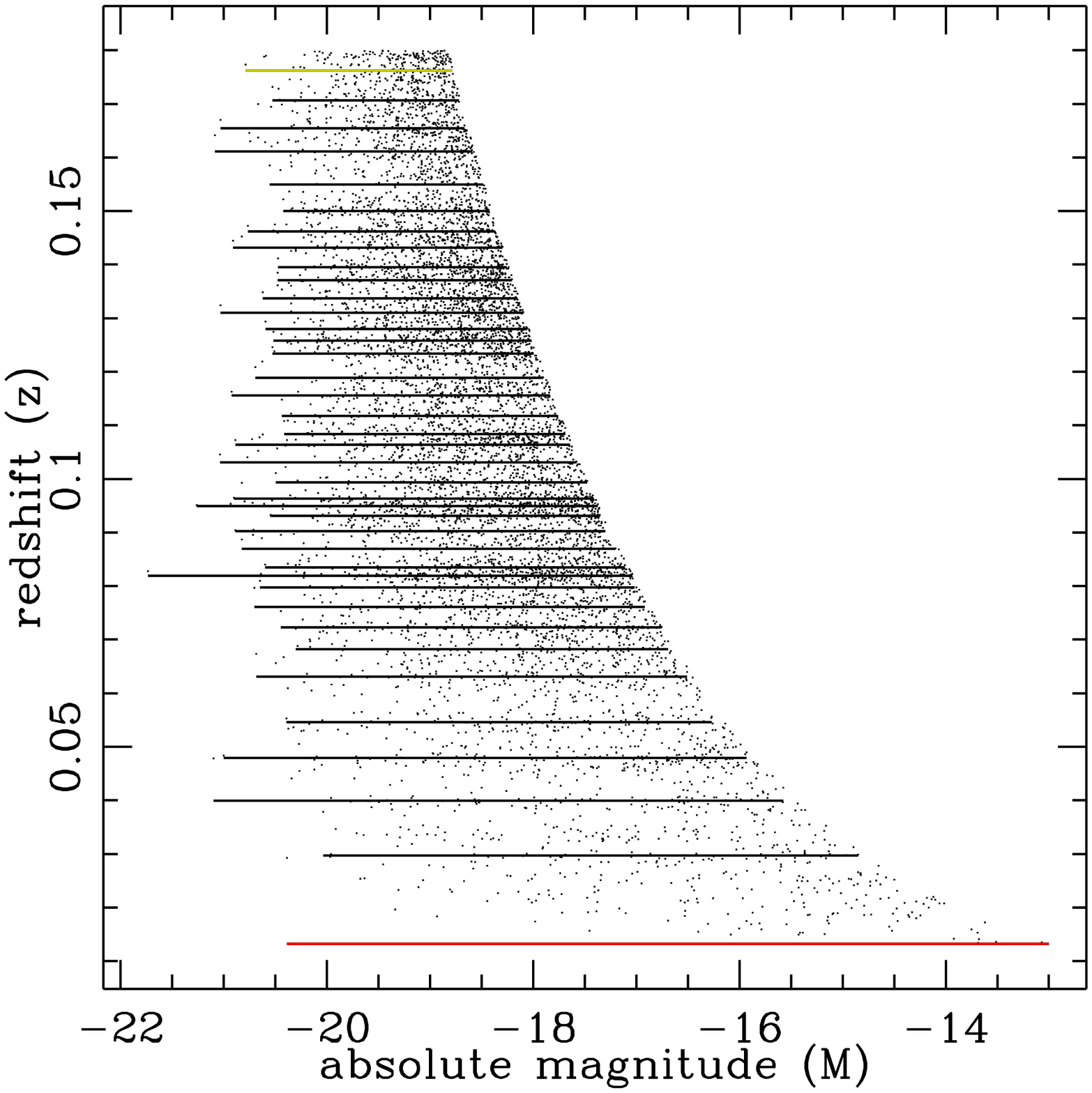}\hfill
		\includegraphics[width=0.46\textwidth]{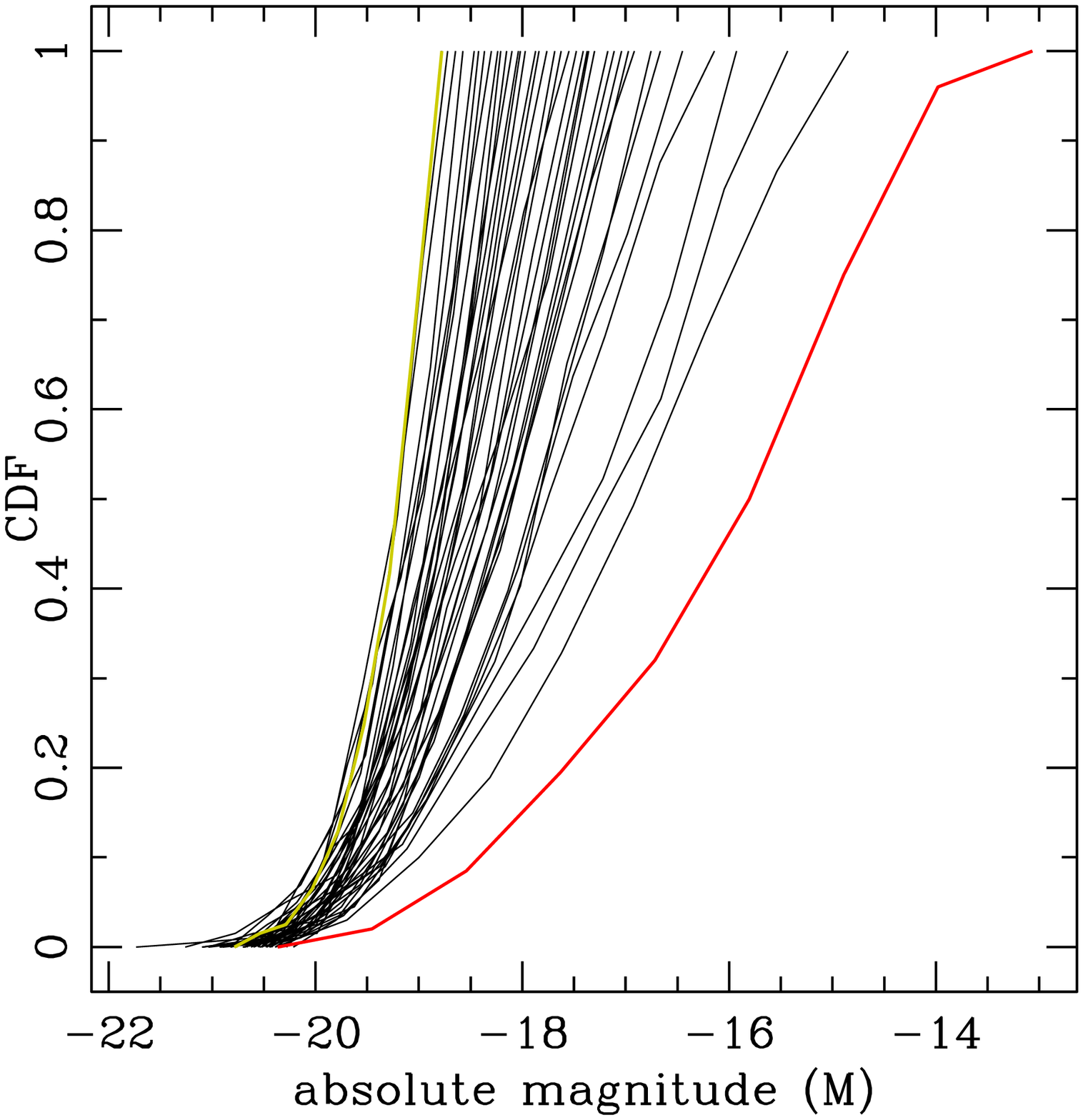}
     \caption{\small Schematic representation of  how we sampled magnitudes to produce our Monte Carlos simulations. The left-hand panel shows the MGC M-z distribution.  The horizontal lines indicate the slices in redshift of equal density within which a CDF of the absolute magnitudes is created. However, it should be noted that the length of each line is merely showing the extent between the brightest and faintest galaxy at that boundary.  The right-hand panel shows the corresponding  CDFs from which a mock magnitude is sampled.  The red line indicates the CDF for the initial redshift bin and the green line, the final bin.
      }
	\label{Fig:z_slice}
  \end{figure*}
\begin{figure}
	\begin{center}
		 \includegraphics[width=0.45\textwidth]{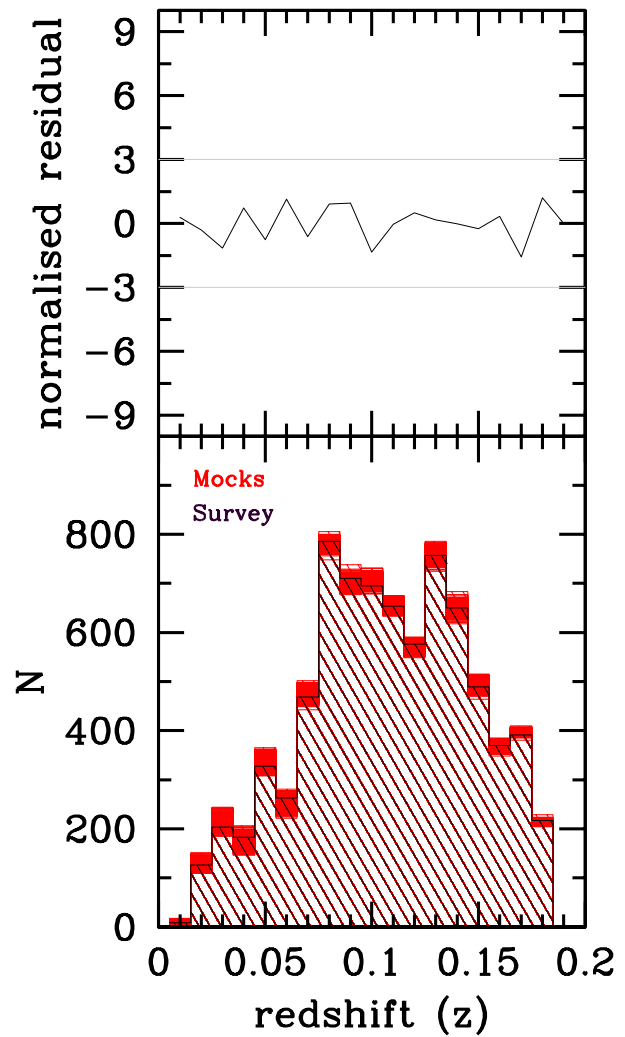}
     \caption{\small The bottom panel shows the MGC redshift distribution (black) compared to a 1000 Monte Carlos (MC) simulated distributions (red).  The simulated redshifts were randomly drawn from the CDF of the observed distribution. However, for consistency with the sampled magnitudes, the density of galaxies equalled that contained within  the z-slices detailed in {\F{Fig:z_slice}}.     The top panel shows the corresponding normalised residuals.
     }
	\label{Fig:z_hist}
	\end{center}
  \end{figure}
\begin{figure}
	\begin{center}
     		 \includegraphics[width=0.5\textwidth]{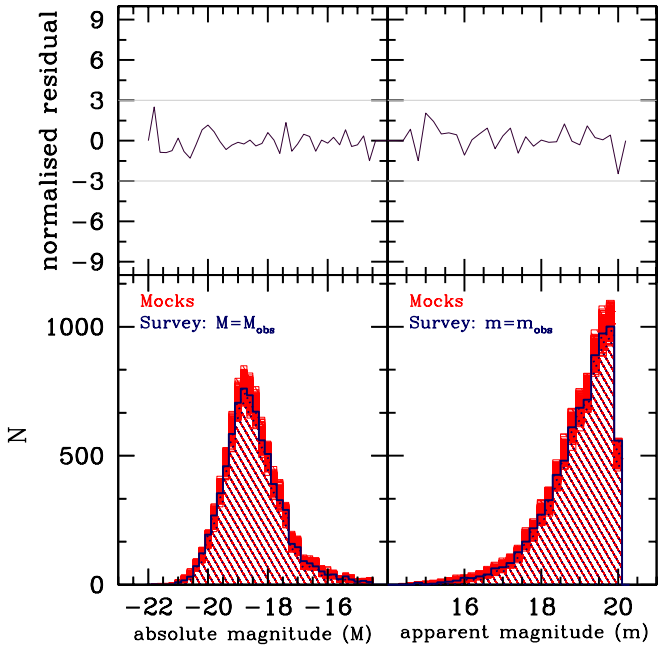}
     \caption{\small The bottom panels show the MGC magnitude distributions (in blue) compared to a 1000 Monte Carlos (MC) simulated distributions (in red).  These MC absolute magnitudes  were drawn using an inverse sampling technique that utilised the MGC survey data (see {\F{Fig:z_slice}}).  The absolute magnitudes were then converted to apparent magnitudes, $m$, as shown on the right-hand panel.  The top panels are the corresponding normalised residuals between the MCs and the  survey data.
      }
	\label{Fig:ammt_hist}
	\end{center}
  \end{figure}
To investigate error propagation for the {\tctv} estimators it was useful to generate realistic Monte Carlo (MC) simulations of a galaxy survey. Clearly there are several approaches to generating such MC simulations.  A popular method is to utilise numerical cosmological simulations \citep*[see e.g.][]{Cole:1998,Norberg:2002b}. This procedure can be summarised  as follows:
\begin{enumerate}
\vspace{-0.2cm}
\item Randomly sample dark matter halos from a cosmological N-body simulation, generated e.g. in a $\Lambda$CDM framework;
\item Employ an algorithm to `assign' galaxies to the sampled dark matter halos \citep*[see e.g.][]{berlind:2002} in a manner that mimics the clustering and the luminosity distribution of the survey under study. Clustering statistics are usually quantified via e.g. the spatial two point correlation function, while luminosities are typically drawn from e.g.  a Schechter function, with parameters, $M_*$, $\alpha$, and $\Phi_*$ which are inferred from observations of the real survey and should represent the present day luminosity function of the target galaxy population;
\item Repeat the above steps as many times as required, selecting the sampled `galaxies' that are consistent with the survey selection function -- at which point an apparent magnitude is generated to be consistent with the redshift of the galaxy derived from the numerical simulation.
\end{enumerate}

In this paper, however, and in the spirit of our desire to develop and apply non-parametric methods, we have employed a rather simpler prescription for generating mock data that retain the same statistical properties as the real survey under study. Specifically, we generate galaxy redshifts and magnitudes by sampling directly from the observed cumulative distribution of these variables in the real galaxy survey\footnote{Note that, in order to simulate accurately the spatial clustering inherent in the real galaxy survey, ideally our mock surveys should {\em also\/} mimic the observed angular coordinates of the real survey.  While this step would be quite straightforward to implement, since in this paper we make no further use of any directional information we do not perform it for the mock galaxy surveys presented here}.  In addition to its robustness, such an approach has the obvious advantage of not requiring the scale of computing power necessary for cosmological N-body simulations -- although one recognised limitation is that the method lacks scope for generating multiple realisations (or sampling from multiple locations in a single realisation) in order to mitigate the impact of cosmic variance -- which may be particularly important in smaller volumes at high redshift \citep*[see e.g.][]{somerville:2004}. Notwithstanding the above limitation, our robust method simulates the {\em observed\/} magnitude and redshift distribution extremely effectively. We now describe the steps involved in our procedure for the specific example of the MGC survey.

We want our mock surveys to reproduce as accurately as possible the observational selection effects to which the real survey data are subject. Of course those selection effects are manifest in a plot of the (uncorrected) $M-Z$ distribution for the MGC galaxies; this is shown in \ref{Fig:z_slice} and we clearly see that the bright and faint apparent magnitude limits render the distribution of absolute magnitude for {\em observable\/} galaxies strongly dependent on redshift.  We explicitly include this dependence as follows:
\begin{enumerate}
\vspace{-0.2cm}
\item We divide up the redshift distribution into a series of redshift bins, each containing an equal number of galaxies.  The boundaries of these bins are indicated by the horizontal lines in the left-hand panel of Figure \ref{Fig:z_slice}.
\item For each redshift bin we compute the sample CDF of absolute magnitude for observable galaxies within that bin.  These sample CDFs are shown in the right-hand panel of Figure \ref{Fig:z_slice}, where we see that they do indeed clearly vary with redshift bin, as expected.  The sample CDFs for the first redshift bin and last redshift bin are denoted by the red and green curves respectively.
\item We also compute the sample CDF of redshift for the entire MGC survey.
\item We then generate redshifts for our mock survey by repeatedly sampling random redshift values from the sample CDF which we computed at step 3, checking as we do so that the number of mock galaxies sampled in each of the bins shown in the left-hand panel of Figure \ref{Fig:z_slice} matches closely the number of {\em real \/} MGC galaxies found in that redshift bin.  We carry out our sampling using the well-known ``inverse sampling" method, based on the probability integral transform, as described in e.g. Section 7.2, Page 287 of Numerical Recipes \citep*{Press:1992}.
\item Finally for each mock galaxy we assign a corresponding absolute magnitude by first identifying to which redshift bin the galaxy belongs and then drawing a random value from the appropriate sample CDF of absolute magnitude which we computed at step 2. Again we use the inverse sampling method to generate random absolute magnitudes.
\end{enumerate}
The bottom panel of  Figure~\ref{Fig:z_hist} shows the resulting redshift histogram for our MGC mocks shown in red, compared to the survey distribution shown in black.  For continuity with our previous studies in this series of papers we have adopted the same redshift limits as in {\CII} where $z_{\min}=0.013$ and  $z_{\max}=0.18$.  As a quality check we show the  normalised residuals, $r$,  for each bin which were computed according to the following relation,
\begin{equation}\label{equ:residual}
r=\frac{{\sum\limits_{i = 1}^n {(\bar z^{\rm MC}_i - z^{\rm MGC}_i)}}}{\sqrt{{\rm Var}(z^{\rm MC}_i )}},
\end{equation}
where $n=1000$ is the number of mock surveys.  $z^{\rm MC}_i$ represents each mock  and $Y$ the MGC survey data for each bin.  Thus, the mean $z^{\rm MC}$ and an unbiased estimate of the variance, ${\rm Var}(z^{\rm mock})$, is given by,
\begin{align}
&\bar z^{\rm MC} = \frac{1}{n}\sum\limits_{i = 1}^n {z^{\rm MC}_i}, \\ \nonumber
{\mbox {and}} \\
&{\rm Var(z^{\rm MC})}=\frac{1}{n-1}{\sum\limits_{i = 1}^n {(z^{\rm MC}_i  - \bar z^{\rm MC})^2 } }.
\end{align}
Assuming that our residuals are Gaussian (which should follow from the central limit theorem) we observe that they lie well within the indicated limits $[-3,3]$, suggesting that our sampled redshifts are consistent with the survey data.

The mock survey absolute magnitudes were assigned with the extra condition imposed such that they must correspond to a galaxy that would be observed within the faint apparent magnitude limit of the survey, {\mlimf}.  Therefore, each sample absolute magnitude was  first generated, then converted to an apparent magnitude, $m$,  via the simple relation,
\begin{equation}\label{equ:absmag}
m_i=M_i+Z_i,
\end{equation}
where $Z_i$ is the distance modulus derived from the simulated redshift, and thus given by,
\begin{equation}
Z_i=5\log(d_{L_i}) +25,
\end{equation}
where in turn $d_{L_{i}}$ is the luminosity distance (in Mpc) of the
i$^{th}$ galaxy, i.e.
\begin{equation}
d_{L_{i}}=(1+z_{i})\left(\frac{c}{H_{0}}\right)\int^{z_{i}}_0
\frac{dz}{\sqrt{(1+z_{i})^{3}\Omega_{m0}+\Omega_{\Lambda0}}},
\end{equation}
in which, to be consistent with {\CII}, we have set the present-day matter density $\Omega_{\rm m0} = 0.3$,  the cosmological constant term $\Omega_{\Lambda0}$ = 0.7, and the Hubble constant $H_{0}$ = 100 kms$^{-1}$ Mpc$^{-1}$. Note that for simplicity we do not correct for $k$- or evolutionary effects, and thus our sampled distribution essentially represents the {\it raw} absolute magnitudes.  This simplification does not affect our ability to probe the error propagation of our completeness estimators.
\begin{figure*}
     		 \includegraphics[width=0.9\textwidth]{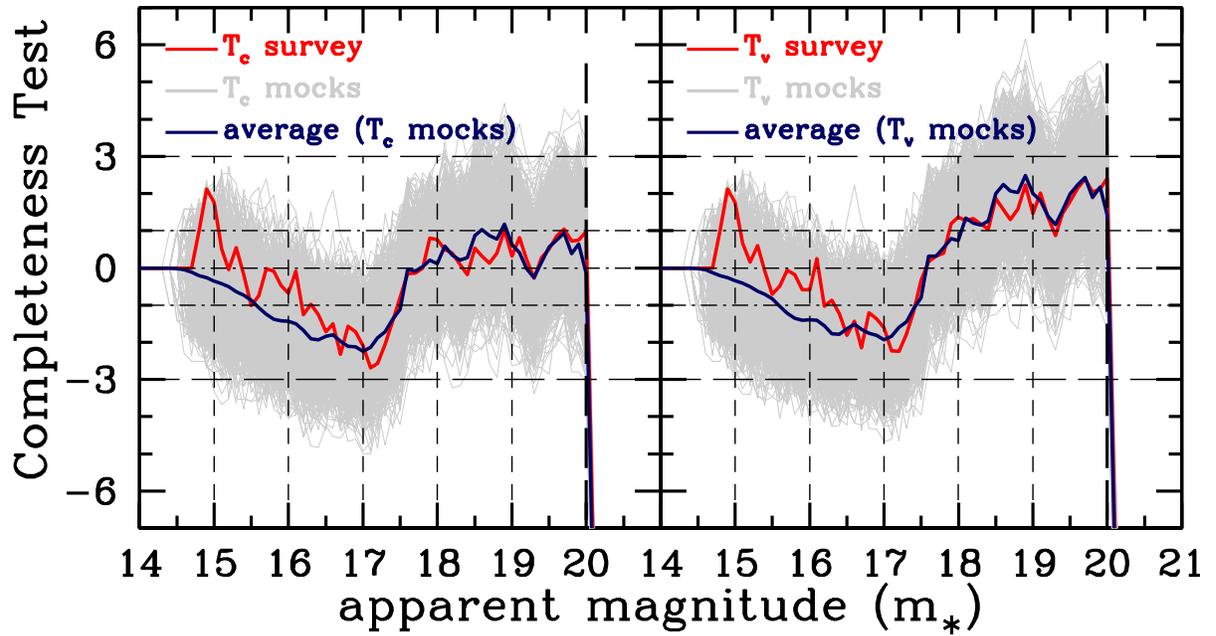}
     \caption{\small {\tctv} statistics applied to the 1000 MGC mocks (grey lines). The completeness results from the actual MGC survey are shown in red and the averaged mocks are shown in blue.  The faint apparent magnitude limit, {\mlimf}, of the MGC survey is indicated by the vertical dashed line at {\mlimf}=20.0~mag.
      }
	\label{Fig:tctv100moks}
  \end{figure*}

The resulting sampled distributions for $M$ and $m$ are shown in red on the  respective bottom left and right panels of Figure~\ref{Fig:ammt_hist}.   As with redshift, we have calculated the normalised residuals for both magnitude distributions; these are shown in the top panels of the figure. We find them to be  within acceptable limits of $[-3,3]$.

Finally, in  {\F{Fig:tctv100moks}}  we apply the  R01 completeness estimators to the mocks and compare them to the completeness results from the actual MGC data originally calculated in {\CII}.  The  {\tctv} superimposed results for the 1000 mock catalogues are shown in grey,  with {\tc} on the left panel and {\tv} on the right panel.  The MGC survey results are shown in red and the averaged {\tctv} values for each {\mstar} for the mocks is indicated by the  blue line.  It is interesting and reassuring to observe  that the averaged {\tctv} curve of the mocks agrees remarkably well with the survey data results.  Of course this offers another method by which one could probe for systematics and/or inconsistencies in any procedure that is adopted to generate Monte Carlo galaxy samples.
\label{lastpage}
\onecolumn
\end{document}